\documentclass{aa}

\usepackage{graphicx}
\usepackage{txfonts}
\usepackage{enumitem}
\usepackage{array}

\usepackage[dvipsnames]{xcolor}

\usepackage{colortbl}
\usepackage{arydshln}
\newcommand{\blockrule}{\arrayrulecolor{black!40}\hdashline[0.4pt/1pt]\arrayrulecolor{black}}
\makeatletter
\renewcommand\@makecaption[2]{%
  \aa@captionfont
  \par\vskip\VSpaceBeforeFigCaption
  \aa@@makecaption{#1}%
  \aa@capstrut
  #2%
  \par
}
\makeatother

\usepackage[
  colorlinks=true,
  linkcolor=blue,
  citecolor=blue,
  urlcolor=blue
]{hyperref}

\begin{document} 
\title{\textsc{stepsic}: A novel initial condition generator for stereographic cosmological simulations}
\author{
    Balázs Pál\inst{1,2}\fnmsep\thanks{pal.balazs@ttk.elte.hu} \and Gábor Rácz\inst{3} \and István Csabai\inst{1} \and István Szapudi\inst{4}
}    
\institute{
    Department of Physics of Complex Systems, ELTE Eötvös Loránd University, Budapest, Hungary
    \and
    Institute for Particle and Nuclear Physics, HUN-REN Wigner Research Centre for Physics, Budapest, Hungary
    \and
    Department of Astronomy, University of Helsinki, Helsinki, Finland
    \and
    Institute for Astronomy, University of Hawaii, 2680 Woodlawn Drive, Honolulu, HI 96822, USA
}
\date{Received ***; accepted ***}

\abstract
{}
{We present \textsc{stepsic}, an open-source initial condition generator for cosmological $N$-body simulations in the periodic cuboid ($\mathbb{T}^{3}$), spherical ($\mathbb{R}^{3}$), and cylindrical ($S^{1}{\times}\mathbb{R}^{2}$) topologies supported by the \textsc{StePS} code. It is the stand-alone successor of the initial condition code used for earlier spherical and cylindrical \textsc{StePS} simulations and extends its construction to periodic cuboids with arbitrary aspect ratios.}
{Gaussian density fields are generated on a periodic Fourier grid, and first- and second-order Lagrangian perturbation theory (LPT) displacements and velocities are interpolated to regular lattices or to relaxed glasses with radially varying particle masses. For the spherical and cylindrical topologies, the fields are computed in a periodic box that encloses the simulated domain, and a multiresolution scheme matches the resolved wavelengths to the local particle mass.}
{For the same white noise, the periodic calculation agrees with the \textsc{monofonic} generator to within $0.05\%$ in the power spectrum and $0.15\%$ in the median particle displacement. On relaxed spherical and cylindrical glasses, particle displacements differ from an independent sum over the retained Fourier modes by the error expected for linear interpolation between mesh cells. With the default enclosing box of one domain diameter, the root-mean-square relative displacement of two points in the high-resolution region differs from its value in the infinite Universe by at most $0.0014\,h^{-1}\mathrm{Mpc}$ out of $0.5\,h^{-1}\mathrm{Mpc}$ at $z=31$. On the scale of the whole domain the box narrows the spread of the mean density between realizations, to a standard deviation of $1.6\times10^{-4}$ instead of $2.0\times10^{-4}$ for a sphere of radius $508\,h^{-1}\mathrm{Mpc}$. A box of two diameters gives $2.0\times10^{-4}$. In one cylindrical realization evolved to $z=0$ from boxes of one and of two diameters, the high-resolution region is displaced as a whole by $0.39\,h^{-1}\mathrm{Mpc}$ between the runs. With this displacement removed, its power spectra differ by at most $1.05\%$. In a cylindrical \textsc{StePS} simulation evolved to $z=0$, first-order initial conditions leave up to $3.6\%$ less nonlinear power than second-order ones, as expected from their transients.}
{}
\maketitle
\nolinenumbers
%

\section{Introduction} \label{sec:intro}
Initial conditions of cosmological simulations are generated based on a specified cosmological model, usually derived from the inflationary paradigm and constrained by observations of the Cosmic Microwave Background~\citep{EisensteinHu1998}. Cosmological initial conditions set the density fluctuations and velocity fields at an early epoch, which then evolve under gravity to form the cosmic web of structures observed today. Conventional cosmological $N$-body simulations employ periodic boundary conditions in a cubic domain, effectively equipping the simulated volume with a $\mathbb{T}^{3} = S^{1}{\times}S^{1}{\times}S^{1}$ (3-torus) topology~\citep{Efstathiou1985,Peebles1980}. This approach has proven extremely successful both for studying structure formation in the standard cosmological setting and for constructing full-sky lightcone catalogs~\citep{Llinares2017,Euclid2025}. However, the $\mathbb{T}^{3}$ topology introduces artificial correlations at scales comparable to the box size, which, through nonlinear mode coupling, can propagate to smaller scales and compromise the simulation even in the regime of interest. Furthermore, the cubic domain is not invariant under the full rotation group $\mathrm{SO}(3)$; consequently, angular momentum is generally not conserved in periodic simulations~\citep{Racz2021}. These limitations make the $\mathbb{T}^{3}$ topology unsuitable for experiments that require non-periodic or observer-centric boundary conditions, or for phenomena that necessitate rotational invariance.

\begin{figure*}[t]
\centering
\includegraphics[width=\textwidth]{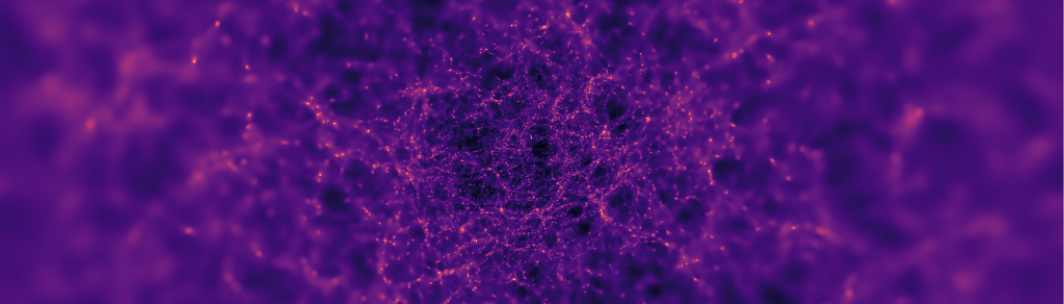}
    \caption{
        Projected density at $z=0$ in the cylindrical $S^{1}{\times}\mathbb{R}^{2}$ \textsc{StePS} simulation initialized with second-order LPT. The image projects the whole domain, which has $R_{\rm 3D}=750\,\mathrm{Mpc}$ and $L_z=200\,\mathrm{Mpc}$. Color intensity represents the projected matter density. The radial decline in mass resolution is visible toward the edges, while filaments, halos, and voids remain resolved in the central region. No color bar is shown because the image is a qualitative view rather than a density measurement. It was rendered with \textsc{py-sphviewer2}~\citep{BenitezLlambay2025} using an orthographic projection at an elevation of $25^\circ$, and the image spans $1350\,\mathrm{Mpc}$ horizontally.
    }
    \label{fig:cylindrical-density}
\end{figure*}

The \textsc{StePS}\footnote{\url{https://github.com/eltevo/StePS}} (STEreographically Projected cosmological Simulations) code~\citep{Racz2019} overcame this limitation by compactifying the $\mathbb{R}^{3}$ manifold into a finite domain via inverse stereographic projection, yielding a spherical volume with non-periodic boundaries~\citep{Racz2018}. This domain is invariant under the full rotation group $\mathrm{SO}(3)$ about its center, which a periodic cube is not. In exchange, the domain gives up translational invariance, because its resolution decreases with distance from the center. More recently, \cite{Racz2026} extended the framework to a cylindrical $S^{1}{\times}\mathbb{R}^{2}$ topology, in which the simulation domain is periodic along a single axis while remaining open in the radial direction. This geometry is appropriate to study the filamentary structure of the Universe and anisotropic cosmological models. In both spherical and cylindrical configurations, the code employs a radially varying particle resolution, with high fidelity near the center and increasingly coarser resolution at large distances, reflecting the geometric compression induced by the projection. Fig.~\ref{fig:cylindrical-density} shows the resulting range of resolved structure in an evolved cylindrical realization. Importantly, all stereographic mapping is performed during the generation of initial conditions, while the \textsc{StePS} $N$-body simulations themselves run entirely in the projected Euclidean space. Established stand-alone codes such as \textsc{2lptic}~\citep{Crocce2006}, \textsc{music}~\citep{HahnAbel2011}, and \textsc{monofonic}~\citep{Michaux2021,Hahn2021} are designed exclusively for fully periodic $\mathbb{T}^{3}$ domains and cannot on their own produce initial conditions for the spherical, cylindrical, or non-cubical volumes that \textsc{StePS} requires. The spherical simulations of \citet{Racz2018,Racz2019} were therefore initialized with \texttt{StePS\_IC}, the initial condition code distributed with \textsc{StePS}, which computes the displacement fields with \textsc{2lptic} or \textsc{n-genic}~\citep{Springel2005} in a periodic box and applies them to the spherical particle load.

Here, we present \textsc{stepsic}\footnote{\url{https://github.com/eltevo/stepsic}}, a stand-alone initial condition generator for \textsc{StePS} cosmological simulations in the $\mathbb{R}^{3}$, $\mathbb{T}^{3}$, and $S^{1}{\times}\mathbb{R}^{2}$ topologies. The code implements Lagrangian perturbation theory (LPT) up to second order, supports several initial particle loads, and includes a multiresolution scheme that matches the displacement fields to the radially varying mass resolution of stereographically projected simulations, akin to the approach of \cite{HahnAbel2011} for zoom-in simulations. The \textsc{stepsic} code is the stand-alone successor of \texttt{StePS\_IC}~\citep[Sec.~3]{Racz2019}, and every \textsc{StePS} simulation in the past was initialized with one of the two codes. The \textsc{stepsic} code keeps the construction of its precursor, which perturbs a relaxed particle load with displacement fields computed on periodic grids whose resolved wavelengths are matched to the local particle mass, but computes these fields and their interpolation to the particles itself instead of taking them from an external periodic generator. An earlier version of \textsc{stepsic} initialized the cylindrical $S^{1}{\times}\mathbb{R}^{2}$ simulation of \citet{Racz2026}. Periodic cuboids with arbitrary aspect ratios are completely new in this release.

This paper is organized as follows. Sec.~\ref{sec:geometries} treats the supported geometries as alternative choices of topology and compares their properties. In Sec.~\ref{sec:particle-load} we describe the initial particle load generation strategies supported by \textsc{stepsic}. The construction of density and velocity fields in Fourier space is detailed in Sec.~\ref{sec:fields}, followed by the Lagrangian perturbation theory implementation in Sec.~\ref{sec:lpt}. Sec.~\ref{sec:interpolation} discusses the interpolation of displacement fields onto particle positions, as well as the multiresolution scheme. Sec.~\ref{sec:validation} presents validation tests, including a comparison with the \textsc{monofonic} code, and Sec.~\ref{sec:performance} reports the computational cost.

\section{Geometries} \label{sec:geometries}
Every cosmological $N$-body simulation replaces the infinite, statistically homogeneous and isotropic Universe by a finite model, and the topology of that model is always a choice with specific consequences (Table~\ref{tab:topologies}). The periodic $\mathbb{T}^{3}$ box preserves translational invariance and uniform resolution, and its Poisson equation can be solved efficiently with Fourier methods. In exchange, it contains no modes longer than the box, fixes the mean density of the box to the cosmic mean, identifies every point with its periodic images, and retains only the discrete rotational symmetry of the box~\citep{Racz2021}. The compactified $\mathbb{R}^{3}$ geometry of \textsc{StePS} makes the opposite trade. It is invariant under all rotations about its center and has no periodic images, but it gives up translational invariance, because its mass resolution decreases with distance from the center~\citep{Racz2018,Racz2019}. The cylindrical $S^{1}{\times}\mathbb{R}^{2}$ geometry is periodic along one axis and open in the other two~\citep{Racz2026}, and an anisotropic cuboid is a $\mathbb{T}^{3}$ topology with direction-dependent fundamental modes. An initial condition generator must therefore produce, for the chosen topology, a Gaussian realization with the statistics of that topology: for $\mathbb{T}^{3}$, those of the discrete Fourier lattice of the box, and for the open geometries, those of the infinite Universe restricted to the simulated domain, sampled on the modes of the periodic box that encloses it (Sects.~\ref{ssec:nonperiodic} and \ref{ssec:embedding}). Differences between topologies, such as the anisotropic large-scale modes of a thin periodic slab, are properties of the chosen model rather than errors of the initial conditions.

\begin{table}
    \centering
    \footnotesize
    \setlength{\tabcolsep}{3pt}
    \renewcommand{\arraystretch}{1.3}
    \begin{tabular*}{\columnwidth}{@{\extracolsep{\fill}}>{\raggedright\arraybackslash}p{2.14cm}>{\raggedright\arraybackslash}p{1.82cm}>{\raggedright\arraybackslash}p{2.03cm}>{\raggedright\arraybackslash}p{2.35cm}@{}}
        \hline\hline
        & $\mathbb{T}^{3}$ cuboid & $\mathbb{R}^{3}$ sphere & $S^{1}{\times}\mathbb{R}^{2}$ cylinder \\
        \hline
        Continuous symmetries & translations & rotations about the center & axial translations and rotations about the axis \\
        \hline
        Rotation group & $D_{2}$ for unequal sides, $O_{h}$ for a cube & $\mathrm{SO}(3)$ & $\mathrm{SO}(2)$ about the axis and half-turns perpendicular to it \\
        \hline
        Boundary & periodic & open, with homogeneous exterior & periodic along the axis, open radially \\
        \hline
        Longest initial modes & box sides & enclosing box & $L_z$ (axial), enclosing box (transverse) \\
        \hline
        Mean density of the domain & fixed & free & free \\
        \hline
        Mass resolution & uniform & decreasing with radius & decreasing with distance from the axis \\
        \hline
    \end{tabular*}
    \caption{Properties of the topologies supported by \textsc{stepsic}. The enclosing box is the periodic Fourier box in which \textsc{stepsic} computes the fields for the open geometries (Sec.~\ref{ssec:nonperiodic}). The mean density of an open domain after displacement is discussed in the same section.}
    \label{tab:topologies}
\end{table}

In \textsc{stepsic}, three distinct simulation geometries are supported that correspond to the capabilities of the \textsc{StePS} $N$-body code: (i) rectangular parallelepiped (cuboid) with arbitrary aspect ratios, (ii) spherical, and (iii) cylindrical domains. The choice of geometry fundamentally determines both the boundary conditions of the simulation and the method of mass normalization required to achieve the correct cosmological density. For cuboids, periodicity can be chosen separately along each axis.

For cuboid $\mathbb{T}^{3}$ geometries, \textsc{stepsic} generates initial conditions within a cuboid of dimensions $(L_x, L_y, L_z)$, with the simulation volume simply calculated as
\begin{equation}
    V_{\rm sim}
    =
    L_x \times L_y \times L_z\,.
\end{equation}
This represents the standard approach used by most cosmological simulation codes with periodic boundary conditions. In $\mathbb{R}^{3}$ spherical geometries, particles are distributed within a sphere of radius $R_{3 \rm D}$, yielding a simulation volume
\begin{equation} \label{eq:volume-sph}
    V_{\rm sim}
    =
    \frac{4}{3}\pi R_{3 \rm D}^3\,.
\end{equation}
In $S^{1} \times \mathbb{R}^{2}$ cylindrical geometries, the simulation domain is a cylinder of radius $R_{3 \rm D}$ whose height equals the extent along the periodic axis $L_{\rm per}$, giving
\begin{equation} \label{eq:volume-cyl}
    V_{\rm sim}
    =
    \pi R_{3 \rm D}^2 \, L_{\rm per}\,.
\end{equation}
By default, \textsc{StePS} applies free boundary conditions in every direction for spherical geometries, while employing periodic boundaries along the axis of the cylinder and free boundaries in the radial directions for cylindrical geometries. In both open geometries, \textsc{StePS} treats the matter outside the simulated domain as homogeneous~\citep{Racz2018,Racz2026}. In all cases, particle masses are automatically rescaled in \textsc{stepsic} to ensure that the mean density of the undisplaced particle load matches the cosmological expectation $\bar{\varrho}_{\rm sim} = \Omega_{\rm m} \varrho_{\rm crit}$, regardless of the chosen geometry. 

\section{Particle load} \label{sec:particle-load}
The \textsc{stepsic} code provides four options to construct an initial particle load: a regular lattice configuration, a random Poisson distribution, concentric shell filling for \textsc{StePS} geometries, and a method to handle externally generated cosmological ``glass'' inputs.

\begin{figure*}
\centering
\includegraphics{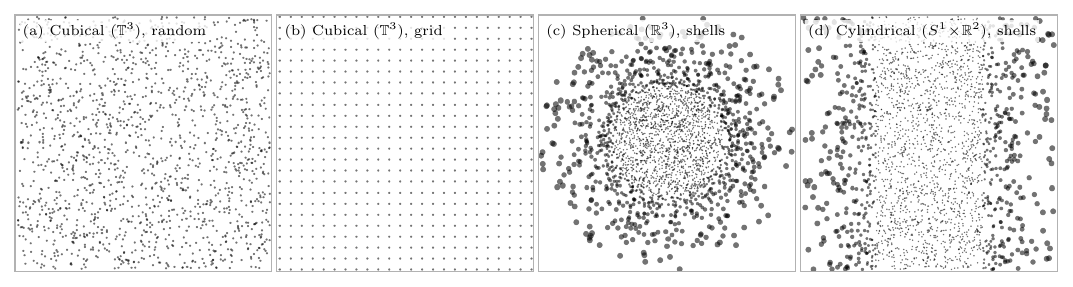}
    \caption{
        Two-dimensional slices through the four particle load configurations supported by \textsc{stepsic}, shown before any Lagrangian displacement is applied or glass relaxation is performed. Each panel displays particles within a thin slab of thickness $25\,h^{-1}\mathrm{Mpc}$ centered on the midplane and oriented perpendicular to the $x$-axis, projected onto the $y$--$z$ plane.
        \textbf{(a)}~Uniform random (Poisson) sampling within a cubical $\mathbb{T}^{3}$ domain.
        \textbf{(b)}~Regular lattice (grid) within the same cubical domain; to preserve the lattice pattern, a single grid layer is shown instead of the finite-thickness slab used in the other panels.
        \textbf{(c)}~Concentric spherical shells for an $\mathbb{R}^{3}$ geometry with free boundary conditions and radially varying mass resolution; marker sizes scale with particle mass to reflect the variable-resolution shell structure.
        \textbf{(d)}~Concentric cylindrical annuli for an $S^{1}{\times}\mathbb{R}^{2}$ geometry, periodic along the vertical axis with free boundaries radially; marker sizes again scale with particle mass.
        In all panels, the full particle set is shown ($N_{\rm part} = 32^{3}$ for the Poisson sample; $N_{\rm grid} = 32^{3}$ for the lattice; $N_{\rm shell} = 1536$ particles per bin, $N_{\rm bins} = 128$ for the shell-based loads, both with $R_{\rm c} = 100\,h^{-1}\mathrm{Mpc}$).
    }
    \label{fig:particle-load}
\end{figure*}

\subsection{Regular lattice (grid)} \label{ssec:grid}
In the most elementary case, particles are placed on a rectilinear mesh with cubic voxels. For a cuboid of side lengths $(L_{x}, L_{y}, L_{z})$, we choose integers $(N_{x}, N_{y}, N_{z})$ such that the voxel spacing is uniform along all three axes. The mesh size $\Delta$ is selected with respect to the shortest side of the simulation box and an arbitrary input $N_{\rm input}$ that defines its resolution as
\begin{equation} \label{eq:cubic-voxel}
    \Delta
    \equiv
    \frac{\min(L_{x}, L_{y}, L_{z})}{N_{\rm input}}\,.    
\end{equation}
Then, the number of grid points along each dimension is set to $N_{i} \approx L_{i} / \Delta$, rounded up to an even integer. Ensuring that voxels are cubical, i.e., the spacing is identical in every Cartesian direction, provides maximal spectral range for the field generation based on the fast Fourier transform (FFT) and serves as a convenient Lagrangian coordinate system for the interpolation of displacement fields. Additionally, the lattice is shifted by half a voxel in each dimension to center the grid within the simulation volume. This particle load type is only available for cubical geometries.

\subsection{Homogeneous Poisson distribution (random)} \label{ssec:random}
A statistically isotropic particle load can be obtained by sampling a homogeneous Poisson point process, in which particle positions are uniformly distributed within the simulation volume $V_{\rm sim}$. For an ideal Poisson realization, the total particle number $N$ is a Poisson-distributed random variable with mean $\langle N \rangle = \lambda |V_{\rm sim}|$, where the intensity $\lambda = \langle n(\mathbf{x}) \rangle = \varrho_{\rm mean}/m_{\rm p}$ defines the target mean number density and $m_{\rm p}$ is the mass of each simulation particle. In practice, \textsc{stepsic} fixes the particle number deterministically to $N \approx \lambda |V_{\rm sim}|$. The $N$ positions are then generated by drawing independent uniform random numbers $(u_{1}, u_{2}, u_{3}) \sim U(0,1)$ and mapping them onto the chosen geometry (cuboid, spherical, or cylindrical) using the appropriate coordinate transformation. This yields an uncorrelated, isotropic particle configuration with no intrinsic structure beyond the Poisson shot noise. Since the resulting shot noise power is typically much larger than the target cosmological power spectrum, this mode is primarily useful for testing purposes and as a starting configuration for glass generation (see below), rather than for producing initial conditions. This particle load type is only available for cubical geometries.

\subsection{Concentric shells} \label{ssec:shells}
For simulations employing the \textsc{StePS} stereographic projection, \textsc{stepsic} can fill the simulation volume with concentric shells (spherical) or annuli (cylindrical), each containing a fixed number of particles $N_{\rm shell}$. The total particle count in this mode is $N_{\rm total} = N_{\rm shell} \times N_{\rm bins}$, where $N_{\rm bins}$ is the number of radial bins. Since the shell volumes generally differ, this gives each bin a distinct particle mass, producing the variable mass resolution intrinsic to \textsc{StePS} simulations~\citep{Racz2018}. This particle load type is only available for \textsc{StePS} geometries, i.e., spherical and cylindrical and should be used as inputs for glass generation using the \textsc{StePS} simulation code.

The radial extent of each bin is determined by one of two binning modes, both of which parametrize the shell edges through the stereographic angle
\begin{equation}
    \omega
    =
    2 \arctan \left( \frac{r}{2 R_{4 \rm D}} \right)\,,
\end{equation}
where $r$ is the comoving radial distance and $R_{4 \rm D}$ is the radius of the compactification sphere. Both methods were adapted from~\cite{Racz2019}.

\begin{enumerate}
    \item \textbf{Constant $\Delta\omega$ binning.} The stereographic angle is divided into equal steps of size
    \begin{equation}
        \Delta\omega
        =
        \frac{\pi}{2\,(N_{\rm bins} + \ell)} \,,
    \end{equation}
    where $\ell$ parametrizes the fractional size of the outermost cell, computed automatically so that the last shell boundary coincides with the simulation radius $R_{3 \rm D}$. The outer edge of the $i$-th bin in the non-compact (Euclidean) space is then
    \begin{equation}
        r_{i}
        =
        R_{4 \rm D} \tan(i \, \Delta\omega) \,.
    \end{equation}
    \item \textbf{Constant volume binning.} Each bin encloses the same volume on the compact 3-sphere $S^{3}$ (or the compact 2-sphere for cylindrical geometries). The bin boundaries are determined by inverting the relation
    \begin{equation}
        f(x)
        =
        x - \sin x \,,
    \end{equation}
    which is monotonically increasing on $[0, 2\pi]$. The maximum stereographic angle is 
    \begin{equation}
        \omega_{\rm max}
        =
        2\arctan \left( \frac{R_{3 \rm D}}{2 R_{4 \rm D}} \right)\,,
    \end{equation}
    and the unit bin size on the compact space is
    \begin{equation}
        \Delta_{\rm bin}
        =
        \frac{2\omega_{\rm max} - \sin(2\omega_{\rm max})}{N_{\rm bins}}\,.
    \end{equation}
\end{enumerate}

Within each spherical shell, particle positions are sampled uniformly in volume. Angular positions are generated as uniformly distributed unit vectors on $S^{2}$ via rejection sampling from the unit cube. Radial positions within the shell $[r_{0}, r_{1}]$---geometrically, a conical frustum---are drawn from the cube-root distribution
\begin{equation} \label{eq:shell-radial-sampling}
    r_{\rm shell}
    =
    \left[ U(0,1) \cdot (r_{1}^{3} - r_{0}^{3}) + r_{0}^{3} \right]^{1/3}\,,
\end{equation}
where $U(0,1)$ is a uniform random variable on the interval $[0,1]$ that ensures uniform volume density. For cylindrical annuli, an analogous square-root sampling
\begin{equation} \label{eq:annulus-radial-sampling}
    r_{\rm annulus}
    =
    \sqrt{U(0,1) \cdot (r_{1}^{2} - r_{0}^{2}) + r_{0}^{2}} \,,
\end{equation}
ensures uniform area density in the $(x, y)$ plane, while the $z$-coordinate is drawn uniformly over $[0, L_z]$. Particle masses in each bin are set to
\begin{equation}
    m_{i}
    =
    \varrho_{\rm mean} \frac{V_{{\rm shell}, i}}{N_{\rm shell}} \,,
\end{equation}
where $V_{{\rm shell}, i}$ is the shell or annulus volume. The mass-weighted centroid of a spherical shell and a cylindrical annulus with uniform density is
\begin{equation}
    \bar{r}_{\rm shell}
    =
    \frac{3}{4} \frac{r_{1}^{4} - r_{0}^{4}}{r_{1}^{3} - r_{0}^{3}} \,,
    \qquad
    \bar{r}_{\rm annulus}
    =
    \frac{2}{3} \frac{r_{1}^{3} - r_{0}^{3}}{r_{1}^{2} - r_{0}^{2}} \,.
\end{equation}
respectively; both are computed in numerically stable factored forms internally.

Since particle masses vary by several orders of magnitude across the simulation volume, particles with very different masses can interact in the high density central region, degrading force accuracy in the $N$-body integrator. To mitigate this, \textsc{stepsic} defines a constant resolution radius $r_{\rm c}$, below which all radial bins are assigned the same particle mass, equal to the mass of the shell whose outer edge coincides with $r_{\rm c}$. Shells beyond $r_{\rm c}$ retain the standard radially increasing masses.
\begin{figure}[t]
\centering
\includegraphics[width=\columnwidth]{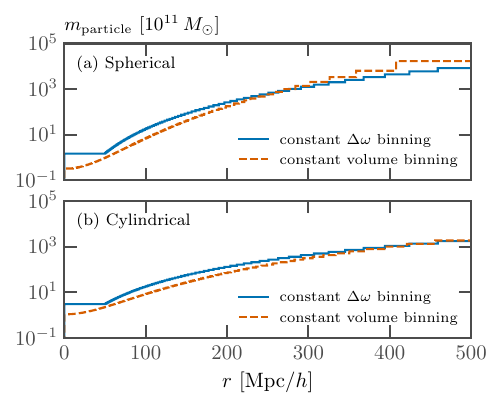}
    \caption{
        Particle mass distribution as a function of radial distance for the constant stereographic angle $\omega$ (solid) and constant compact-space volume (dashed) binning modes in the spherical $\mathbb{R}^{3}$ geometry (top) and the cylindrical $S^{1}{\times}\mathbb{R}^{2}$ geometry along the non-periodic $\mathbb{R}^{2}$ directions (bottom). The constant $\Delta\omega$ profile includes a constant-resolution inner zone with $r_{\rm c} = 50\,h^{-1}\mathrm{Mpc}$. The compactification diameter is $D_{4\mathrm{D}} = 75\,h^{-1}\mathrm{Mpc}$, the simulation radius $R_{3\mathrm{D}} = 500\,h^{-1}\mathrm{Mpc}$, both modes use $N_{\rm bins} = 224$ radial bins with $N_{\rm shell} = 12\,288$ particles per shell, and the cylindrical domain has a periodic extent of $L_{z} = 200\,h^{-1}\mathrm{Mpc}$.
    }
    \label{fig:shell-mass}
\end{figure}
The two binning modes produce qualitatively different mass profiles, whose concrete behavior is illustrated in Fig.~\ref{fig:shell-mass} for the configuration specified in its caption. In the constant $\Delta\omega$ scheme, the transition from the flat inner zone to the steeply rising outer region is gradual; for the parameters shown, this transition spans roughly one decade in mass over $\sim 100\,h^{-1}\mathrm{Mpc}$. The constant volume scheme lacks this plateau by construction and rises more steeply at intermediate radii, but the two profiles converge at large radii where the outermost shells enclose comparable physical volumes regardless of the binning scheme (beyond $r \gtrsim 300\,h^{-1}\mathrm{Mpc}$ in the example of Fig.~\ref{fig:shell-mass}). The cylindrical geometry yields systematically lower masses at all radii than the spherical case, as expected from the different volume scaling of the corresponding shells as discussed in Eqs.~\eqref{eq:volume-sph} and \eqref{eq:volume-cyl}. This hierarchy holds independent of the specific parameter choices. For the configuration shown, the total dynamic (mass) range exceeds four orders of magnitude, which is a direct consequence of the stereographic compactification and grows with increasing $R_{3\mathrm{D}}/D_{4\mathrm{D}}$. The innermost shells have particle masses of order $\sim 10^{10}\,M_{\odot}$, sufficient to resolve Milky Way-mass halos with $\mathcal{O}(10^{2})$ particles, while the simulation volume extends to $R_{3\mathrm{D}} = 500\,h^{-1}\mathrm{Mpc}$ with only $N_{\rm total} \sim 2.75 \times 10^{6}$ particles in total.

\subsection{Glass.} \label{ssec:glass}
A ``glass'' configuration is obtained by evolving a uniform random sample under repulsive gravity until the Fourier-space power is suppressed well below the Poisson shot-noise level, producing a nearly uniform particle distribution that lacks the lattice symmetries of a grid.\footnote{Historically, glass loads were introduced to reduce anisotropic artifacts present in grid initial conditions at early timesteps~\citep{White1994}.} The \textsc{stepsic} code reads such configurations generated by external codes, including the \textsc{gadget} suite~\citep{Springel2021} and \textsc{StePS}. Glasses for the spherical and cylindrical geometries are made by \textsc{StePS} itself~\citep[Sec.~3.1]{Racz2019}. Starting from the concentric-shell load of Sec.~\ref{ssec:shells}, it integrates the equations of motion with reversed gravity in an Einstein--de Sitter background, including the force of the homogeneous exterior, while every particle keeps the mass of its shell.

The \textsc{stepsic} code multiplies all coordinates of an input glass by one factor, so that the glass fills the configured domain, and then rescales the masses to the target mean density. For a spherical or cylindrical glass the factor is $R_{3\mathrm{D}}/R^{\rm glass}$, with $R^{\rm glass}$ the radius recorded in the glass file. For a cuboid it is $L_z/L_z^{\rm glass}$. This similarity transformation multiplies every pairwise gravitational force by the same factor, so a relaxed glass remains relaxed. A cylindrical glass keeps the ratio of its period to its radius, so the configured period must equal $L_z^{\rm glass}R_{3\mathrm{D}}/R^{\rm glass}$. The \textsc{stepsic} code never stretches a glass anisotropically, because that would change the force balance and require a new relaxation. A cuboid or cylinder with a different aspect ratio needs a glass relaxed in that shape.

\section{Fields} \label{sec:fields}

Before perturbations can be applied through Lagrangian perturbation theory, one must first construct the underlying fluctuation field that encodes the statistical properties of the cosmological model. In practice, this requires generating a Gaussian random overdensity field $\delta(\mathbf{k})$ in Fourier space whose two-point statistics reproduce the desired linear matter power spectrum at some initial cosmological redshift $z_{\rm init}$, representing the early Universe and the starting point of the simulation. This step provides the input from which displacements and velocities are subsequently derived.

\subsection{Fourier grid.}
A natural framework for constructing an overdensity field is Fourier space. The reason is twofold. First, the primordial density perturbations are assumed to form a zero-mean (centered) Gaussian random field~\citep{Sirko2005}, which is fully characterized by its two-point correlation function,
\begin{equation} \label{eq:lss-two-point-cf}
    \xi(\mathbf{r})
    =
    \frac{1}{(2\pi)^3}\int\operatorname{d}^3\!k\,P(k)\,e^{i\mathbf{k}\cdot\mathbf{r}}\,.
\end{equation}
Since the perturbations are statistically isotropic and homogeneous, $\xi$ depends only on the separation magnitude $r = |\mathbf{r}|$, and the three-dimensional Fourier integral in Eq.~\eqref{eq:lss-two-point-cf} reduces to a one-dimensional relation between $\xi(r)$ and the matter power spectrum $P(k)$ through the zeroth-order spherical Bessel transform~\citep{Peebles1980}:
\begin{equation} \label{eq:lss-pk-xi}
    \xi(r)
    =
    \frac{1}{2\pi^2}\int_{0}^{\infty}\!\operatorname{d}k\,k^2\,P(k)\,
    \frac{\sin(kr)}{kr}\,,
\end{equation}
where $k \equiv |\mathbf{k}|$ throughout this paper. The linear matter power spectrum has the form
\begin{equation} \label{eq:lss-pk}
    P(k,z)
    \propto
    A_s\,k \left( k/k_{\rm p} \right)^{\,n_{\rm s}-1} T^2(k)\,D_{1,(+)}^{2}(z)\,,
\end{equation}
which specifies the variance of each Fourier mode independently; thus the statistics of $\delta(\mathbf{k})$ are diagonal in Fourier space. Second, calculations on a Fourier grid are particularly efficient: fast Fourier transforms provide a direct mapping between physical and Fourier space, and the Hermitian symmetry of the Fourier modes guarantees that the inverse transform yields a real-valued density field $\delta(\mathbf{x})$.

The discrete Fourier grid is constructed with the same isotropic voxel spacing $\Delta$ as the cubical lattice particle load defined in Eq.~\eqref{eq:cubic-voxel}. This defines a rectangular mesh in Fourier space with fundamental frequencies
\begin{equation}
    k_{{\rm f},\alpha} = \frac{2\pi}{L_{\alpha}}
\end{equation}
along each axis. Each grid point is associated with a wavevector
\begin{equation}
    \mathbf{k}
    =
    (k_{x}, k_{y}, k_{z})
    =
    2\pi
    \left(
        \frac{n_{x}}{L_{x}},
        \frac{n_{y}}{L_{y}},
        \frac{n_{z}}{L_{z}}
    \right),
\end{equation}
where $N_{\alpha}$ denotes the number of mesh points along axis $\alpha$, and the corresponding integer mode indices satisfy
\begin{equation}
    n_{\alpha} \in \left[-\frac{N_{\alpha}}{2}, \frac{N_{\alpha}}{2}\right)
\end{equation}
in the standard FFT ordering. The wavevector magnitude is then
\begin{equation}
    |\mathbf{k}|
    \equiv
    k
    =
    \sqrt{k_{x}^2 + k_{y}^2 + k_{z}^2}
    =
    2\pi
    \sqrt{
        \left( \frac{n_{x}}{L_{x}} \right)^2
        +
        \left( \frac{n_{y}}{L_{y}} \right)^2
        +
        \left( \frac{n_{z}}{L_{z}} \right)^2
    }\,.
\end{equation}
When the simulation domain is anisotropic (e.g., $L_{x} \neq L_{y} \neq L_{z}$) and periodic boundary conditions are imposed, the admissible Fourier modes lie on a rectangular lattice with direction-dependent fundamental spacings $k_{{\rm f},\alpha}$. Since the target input spectrum $P(k)$ is isotropic, the finite set of represented modes is then not isotropic in the strict sense. In particular, long-wavelength modes are truncated anisotropically, which leads to direction-dependent variances in derived quantities such as the displacement field $\boldsymbol{\Psi}$ which maps each particle from its initial (Lagrangian) position to its perturbed position and whose detailed construction is presented in Sec.~\ref{sec:lpt}.

Two distinctions matter for the interpretation of this construction. First, a periodic cuboid is a $\mathbb{T}^{3}$ topology in its own right, not a subvolume of an isotropic Universe (Sec.~\ref{sec:geometries}). A Gaussian realization on it contains exactly the modes of its anisotropic lattice, which suppresses the displacement variance along short box axes. Cutting a region of the same shape from a larger periodic box instead retains the statistics of the larger box, but the region is not periodic across its faces. The two constructions describe different models, and Sec.~\ref{sec:validation} compares them quantitatively. Second, this geometric effect differs from numerical resolution effects. By scaling the mesh dimensions with the box dimensions such that the physical voxel size $\Delta$ is identical along all axes, \textsc{stepsic} removes spurious anisotropies associated with anisotropic resolution. However, the infrared cutoff remains set by the box dimensions $L_{\alpha}$, so anisotropic large-scale modes persist whenever $L_{x}$, $L_{y}$, and $L_{z}$ differ.

\begin{figure}[t]
\centering
\includegraphics[width=\columnwidth]{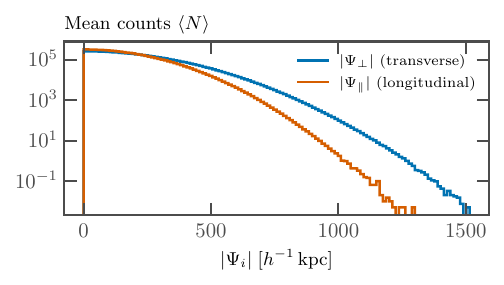}
    \caption{
        Distribution of per-component absolute displacements in a slab geometry with dimensions $L_{x} = L_{y} = 1000\,h^{-1}\mathrm{Mpc}$ and $L_{z} = 200\,h^{-1}\mathrm{Mpc}$, averaged over $200$ realizations using second-order LPT at $z=31$. The transverse curve $|\Psi_\perp|$ is the mean of the $x$ and $y$ components, exploiting the exact $x \leftrightarrow y$ symmetry of the slab. The longitudinal component $|\Psi_\parallel| \equiv |\Psi_z|$ is strongly suppressed, reflecting the anisotropic infrared cutoff imposed by the box geometry. The fundamental mode along $z$ is $k_{{\rm f},z} = 2\pi/L_z$, five times coarser than $k_{{\rm f},x} = k_{{\rm f},y} = 2\pi/L_x$, so large-scale power that would contribute to $\Psi_z$ in a cubic domain is absent from the discrete mode lattice. The measured variance ratio in the simulation ensemble above was $\langle |\Psi_\perp|^2 \rangle / \langle |\Psi_\parallel|^2 \rangle \approx 1.54$.
    }
    \label{fig:validation-slab}
\end{figure}

This effect is illustrated in Fig.~\ref{fig:validation-slab},
where we show the distribution of per-component absolute displacements
for a $5{:}1$ slab geometry averaged over $200$ independent realizations. The transverse displacement $|\Psi_\perp| \equiv \tfrac{1}{2}(|\Psi_x| + |\Psi_y|)$ exploits the exact $L_x = L_y$ symmetry of the domain; we average the two components because their individual histograms are dominated by the pair of fundamental modes $\mathbf{n} = (\pm 1, 0, 0)$ and $(0, \pm 1, 0)$, whose independent stochastic amplitudes induce an $x$\,--\,$y$ variance ratio that follows an $F(2,2)$ distribution and converges only as $\mathcal{O}(N_{\rm real}^{-1/2})$ with respect to a small effective number of modes. The longitudinal component $|\Psi_\parallel| \equiv |\Psi_z|$ is strongly suppressed relative to the transverse direction, consistent with the missing large-scale $z$-modes discussed above. The fundamental mode along $z$ is $k_{{\rm f},z} = 2\pi / L_z$, five times coarser than $k_{{\rm f},x} = k_{{\rm f},y} = 2\pi / L_x$, so large-scale power that would contribute to $\Psi_z$ in a cubic domain is absent from the discrete mode lattice. For a cubic box of the same resolution, all three displacement components are statistically indistinguishable.

\subsection{White noise field.} \label{ssec:white-noise}
The initial seed field is a Gaussian random field $\eta(\mathbf{k})$ with zero mean and unit variance on the discrete Fourier lattice defined above. It is drawn with the counter-based Philox4x64-10 generator~\citep{Salmon2011}, whose key is the random seed and whose counter is the integer wave vector $(n_x,n_y,n_z)$ of the mode. The value of a mode therefore depends only on the seed and on its wave vector, not on the size of the mesh. The multiresolution scheme of Sec.~\ref{ssec:multires} relies on this property. The mean density contrast of the periodic box vanishes when the zero-frequency mode is set to zero,
\begin{equation}
    \eta(\mathbf{k}=\mathbf{0}) = 0\,,
\end{equation}
and the Nyquist planes of each mesh are also set to zero. The resulting field is a realization of an uncorrelated Gaussian white noise process on the discrete mode lattice. The white noise realization can be saved to disk, which allows direct cross-comparison with other initial condition generators using the identical random field.

\subsection{Overdensity field.}
To obtain the correct statistics of matter fluctuations, the white noise amplitudes are rescaled according to the target linear power spectrum $P(k, z_{\rm init})$ at the initial simulation redshift $z_{\rm init} \equiv z$. The \textsc{stepsic} code computes the present-day linear power spectrum of total matter (cold dark matter, baryons, and massive neutrinos), $P(k,z{=}0)$, with \textsc{camb}~\citep{Lewis2000}. Optionally, the primordial amplitude $A_{\rm s}$ is rescaled so that this spectrum has the requested $\sigma_8$. The spectrum is then back-scaled to the initial redshift with the square of the linear growth factor,
\begin{equation}
    P(k,z)
    =
    P(k,z{=}0)\,D_{1,(+)}^{2}(z)\,.
\end{equation}
The growth factor $D_{1,(+)}(z)$ is calculated by the \textsc{colossus}~\citep{Diemer2018} library using the cosmological parameters specified in a run of \textsc{stepsic}. This back-scaling procedure is the default when the power spectrum is computed internally by \textsc{camb}. Alternatively, a pre-computed spectrum $P(k, z_{\rm init})$ may be supplied directly, in which case no growth-factor rescaling is applied. This mode is useful when the $N$-body code e.g., incorporates general-relativistic corrections that account for the late-time evolution of the transfer function, or when the growth function is non-trivial, e.g., in $w$CDM or Chevallier--Polarski--Linder (CPL) dark-energy parametrizations. Following \citet{Sirko2005}, \citet{BaglaPadmanabhan1997}, and \citet{KlypinHoltzman1997}, this power spectrum is interpolated in $\log k$--$\log P(k)$ space onto the discrete wavevector magnitudes $|\mathbf{k}|$ of the Fourier grid, and the overdensity modes are constructed as
\begin{equation} \label{eq:delta-k-gaussian}
    \delta(\mathbf{k},z)
    =
    \eta(\mathbf{k}) \sqrt{\frac{P(|\mathbf{k}|,z)}{\Delta^{3}}}\,,
\end{equation}
where $\eta(\mathbf{k})$ is the white noise realization and $\Delta^3$ is the physical volume of a single voxel.

Eq.~\eqref{eq:delta-k-gaussian} yields a standard Gaussian realization in which both the phases and the amplitudes of the Fourier modes are drawn randomly, consistent with the predictions of the standard $\Lambda$CDM model. The \textsc{stepsic} code additionally supports two variance-reduction techniques:
\begin{enumerate}[label=(\roman*)]
    \item \textbf{Paired realizations.} The pair of a realization inverts the sign of every mode, $\delta(\mathbf{k}) \to -\delta(\mathbf{k})$, and hence of the whole field, $\delta(\mathbf{x}) \to -\delta(\mathbf{x})$. This is the phase shift by $\pi$ of \citet{AnguloPontzen2016}. No other global phase factor preserves the Hermitian symmetry that keeps $\delta(\mathbf{x})$ real. Cosmic variance partially cancels when the two runs are averaged~\citep{Pontzen2016}.
    \item \textbf{Fixed-amplitude realizations.} The Rayleigh-distributed amplitudes $|\eta(\mathbf{k})|$ are set to unity, so that every mode has the amplitude implied by $P(|\mathbf{k}|)$ and only the phases are random. Combined with the paired technique above (``paired-and-fixed''), this further suppresses cosmic variance~\citep{AnguloPontzen2016,Klypin2020} and is the configuration used for the power-spectrum recovery tests of Sec.~\ref{ssec:periodic-boxes}. We note that fixing amplitudes artificially reduces the power spectrum variance and therefore biases higher-order statistics such as the covariance. Therefore, fixed-amplitude realizations should only be used for first-order statistical comparisons.
\end{enumerate}

As for the white noise field, $\delta(\mathbf{k},z)$ can also be saved to disk. Finally, the inverse FFT of $\delta(\mathbf{k},z)$ yields the physical overdensity field $\delta(\mathbf{x},z)$.

\subsection{Non-periodic particle domains.} \label{ssec:nonperiodic}
For the open geometries, the periodic box of the Fourier grid is just a sampling device, and it is independent of the simulated topology. For a spherical domain of radius $R_{3\rm D}$ inside a periodic box of side length $L$, we define the padding ratio $\alpha=L/(2R_{3\rm D})$. For a cylinder, the transverse ratios are $\alpha_x=L_x/(2R_{3\rm D})$ and $\alpha_y=L_y/(2R_{3\rm D})$, and we use $L_x=L_y$. The axial length $L_z$ is the physical ``height'' of the cylinder along its periodic axis. Unless stated otherwise, we use $\alpha=1$, as in earlier \textsc{StePS} simulations~\citep{Racz2018}, and \citet[Sec.~3.2]{Racz2019} discussed the choice of $L$ relative to $R_{3\rm D}$.

The first-order Poisson equation, the quadratic source of the second-order displacement, and the interpolation to the particles are all evaluated in this periodic box. With $\alpha=1$, the initial field therefore inherits two of its properties: (i) it contains no modes longer than the box, and (ii) it is correlated with its periodic images~\citep{Sirko2005,HahnAbel2011}. However, the periodicity does not extend to the dynamics. \textsc{StePS} evolves the particles with open boundaries (sphere) or with periodicity along the axis only (cylinder), and with a homogeneous exterior. In a $\mathbb{T}^{3}$ simulation, by contrast, periodic forces act throughout the evolution, and \citet{Racz2021} found that they bias the power spectrum of small boxes as much as the missing long-wavelength modes do. The box thus acts on the initial field alone, and it does so in two ways: on pairs of points, through their covariance, and on the domain as a whole, through its mean density and its mean shear. We describe the two in turn.

Both inherited properties change the two-point statistics of the field, which describe a Gaussian field completely. We consider two fields $a$ and $b$ that are linear in the density contrast, $a(\mathbf{k})=A(\mathbf{k})\,\delta(\mathbf{k})$ and $b(\mathbf{k})=B(\mathbf{k})\,\delta(\mathbf{k})$, with transfer functions $A$ and $B$. Their covariance is
\begin{equation} \label{eq:covariance-definition}
    C_{AB}(\mathbf{r})
    =
    \langle a(\mathbf{x}+\mathbf{r})\,b^*(\mathbf{x})\rangle\,.
\end{equation}
In the periodic box it is
\begin{equation} \label{eq:periodic-covariance}
    \begin{aligned}
        C_{AB}^{\rm per}(\mathbf{r})
        &=
        \frac{1}{V}\sum_{\mathbf{k}\in\mathcal{K}\setminus\{\mathbf{0}\}}
        P(k)\,|F(\mathbf{k})|^2 A(\mathbf{k})B^*(\mathbf{k})
        e^{i\mathbf{k}\cdot\mathbf{r}} \\
        &\xrightarrow[V\to\infty]{\Delta\ {\rm fixed}}
        \int_{\mathcal{K}_\Delta}\frac{\operatorname{d}^3k}{(2\pi)^3}
        P(k)\,|F(\mathbf{k})|^2 A(\mathbf{k})B^*(\mathbf{k})
        e^{i\mathbf{k}\cdot\mathbf{r}}\,.
    \end{aligned}
\end{equation}
Here $\mathcal{K}$ contains the retained discrete modes, and $\mathcal{K}_\Delta$ is bounded by the Nyquist frequencies at fixed cell spacing $\Delta$. A diagnostic may impose an additional Fourier filter $F$. Otherwise, $F=1$. With the Fourier convention
\begin{equation} \label{eq:fourier-convention}
    \delta(\mathbf{x})
    =
    \frac{1}{V}\sum_{\mathbf{k}}\delta(\mathbf{k})\,e^{i\mathbf{k}\cdot\mathbf{x}}\,,
\end{equation}
the transfer function is $A=1$ for density and $A_i=i k_i/k^2$ for a displacement component. Replacing $\mathcal{K}_\Delta$ by all wavenumbers is justified only after the ultraviolet contribution has been shown to be negligible. In the cylindrical limit, the transverse sums become integrals but $k_z=2\pi n_z/L_z$ remains discrete.

The covariance of two points depends only on their separation vector, so the box does not single out any part of the domain. The box acts on pairs of points according to how far apart they are. For separations much smaller than $L$, only property (i) matters. The sum differs from the integral by the waves longer than the box, which change little across such a pair. For separations comparable to $L$, property (ii) matters as well. A point displaced by $L$ along a box axis is the same point, so at $\alpha=1$ the two points where a box axis crosses the boundary of the domain receive identical displacements. Section~\ref{ssec:embedding} evaluates Eq.~\eqref{eq:periodic-covariance} for boxes with $\alpha=1$, 2, and 3 and for the infinite Universe, and gives the reasons for the default.

Integrated over the domain, the same covariance gives the variance of the mean density and of the mean shear of the domain as a whole. Here a third property of the periodic box, its vanishing mean density contrast, does not carry over to the domain. Setting the zero mode to zero (Sec.~\ref{ssec:white-noise}) fixes the mean of the box, but not the mean within the selected sphere or cylinder! In a periodic box, the displacement field has no net flux through the box boundary, hence the mean density of the box is conserved. In contrast, in an open domain the net flux does not vanish, and after displacement, the particles occupy a volume that differs from $V$ by $\oint_{\partial V}\boldsymbol{\Psi}\cdot\operatorname{d}\mathbf{S}$. Counting the homogeneous exterior that \textsc{StePS} assumes, the mean density contrast within the original boundary $\partial V$ is then, to first order,
\begin{equation} \label{eq:monopole}
    \bar{\delta}_V
    =
    -\frac{1}{V}\oint_{\partial V}\boldsymbol{\Psi}\cdot\operatorname{d}\mathbf{S}
    =
    \frac{1}{V}\int_{V}\delta(\mathbf{q})\,\operatorname{d}^3q\,,
\end{equation}
the average of the linear density contrast over the domain. The same boundary integral with both indices free gives the mean strain of the domain,
\begin{equation} \label{eq:mean-strain}
    \bar{E}_{ij}
    =
    \frac{1}{V}\int_{V}\partial_i\Psi_j\,\operatorname{d}^3q
    =
    \frac{1}{V}\oint_{\partial V} n_i\Psi_j\,\operatorname{d}S\,,
\end{equation}
whose trace is $-\bar{\delta}_V$, the monopole with a negative sign, and whose traceless part $\bar{T}_{ij}$ is the mean shear of the domain. For the cylinder the indices run over the two transverse axes. In an isotropic Universe the off-diagonal components carry three fifths of the variance of $\bar{T}_{ij}$ for the sphere, and one half for the cylinder.

The \textsc{stepsic} code keeps the monopole $\bar{\delta}_V$, so it neither subtracts nor compensates it. The particle masses fix the total mass of the load to $\bar{\varrho}_{\rm sim}V_{\rm sim}$ (Sec.~\ref{sec:geometries}), but not the mass within $\partial V$ after displacement. The resulting $\bar{\delta}_V$ is a physical property of a finite region of the Universe, which a periodic box cannot represent. However, \textsc{StePS} evolves it self-consistently. Since the exterior is homogeneous, an overdense domain evolves like a region of higher mean density~\citep{Racz2018}. At the initial epoch $\bar{\delta}_V$ is small. For a sphere of radius $508\,h^{-1}\mathrm{Mpc}$, the linear top-hat standard deviation of $\bar{\delta}_V$ is $2.0\times10^{-4}$ at $z=31$ in the infinite Universe and $1.6\times10^{-4}$ in a box with $\alpha=1$, which contains no waves longer than the domain. Table~\ref{tab:embedding} in Sec.~\ref{ssec:embedding} compares these predictions, and those for the mean shear, with generated initial conditions.

\section{Lagrangian perturbation theory} \label{sec:lpt}
The final step in traditional cosmological initial condition generation is to calculate the displacement and velocity fields of particles. This is modeled using Lagrangian perturbation theory (LPT), which describes how mass elements move from their initial Lagrangian positions $\mathbf{q}$ to their final Eulerian positions $\mathbf{x}(\mathbf{q},z)$ under gravitational evolution~\citep{Bernardeau2002}. The \textsc{stepsic} implementation follows the conventions established by \textsc{2lptic}~\citep{Crocce2006} and \textsc{monofonic}~\citep{Hahn2021}, ensuring compatibility with standard cosmological simulation workflows while extending support to the non-standard geometries required by \textsc{StePS}.

In the Lagrangian framework, the trajectory of a mass element is expressed as a perturbative expansion up to the $n$-th order as
\begin{equation}
    \mathbf{x}^{(n)}(\mathbf{q},\tau)
    =
    \mathbf{q}
    +
    \mathbf{\Psi}^{(1)}(\mathbf{q},\tau)
    +
    \mathbf{\Psi}^{(2)}(\mathbf{q},\tau)
    +
    \dots + \mathbf{\Psi}^{(n)}(\mathbf{q},\tau)\,,
\end{equation}
where $\tau$ denotes conformal time and $\mathbf{\Psi}^{(i)}$ represents the $i$-th order displacement field. In its current form, the \textsc{stepsic} code implements both the first-order (1LPT, Zel'dovich approximation) and the second-order (2LPT) solutions\footnote{Further extensions to higher-order LPT are considered by e.g., \cite{Hahn2021} or \cite{Hahn2024}, however, their findings indicate that the improvements are negligible beyond the third order for typical cosmological applications. An extension to third-order LPT is planned for a future release of \textsc{stepsic}.}. There are two equivalent conventions for expressing the time dependence of the displacement fields.
\begin{enumerate}[label=(\roman*)]
    \item \textbf{Direct scaling form.} 
    \begin{equation}
        \mathbf{x}^{(n)}(\mathbf{q},z)
        =
        \mathbf{q}
        +
        D_{1}^{({+})}(z)\,\mathbf{\Psi}^{(1)}(\mathbf{q})
        +
        D_{2}^{({+})}(z)\,\mathbf{\Psi}^{(2)}(\mathbf{q})
        + \cdots
    \end{equation}
    \item \textbf{Backscaled form.} The linear power spectrum $P(k,0)$ is rescaled to the initial redshift $z$ and the displacement fields are constructed at that redshift using $\delta(\mathbf{k},z)$, effectively containing the growth factors implicitly. In this case, the particle positions at $z$ are given by
    \begin{equation}
        \mathbf{x}^{(n)}(\mathbf{q},z)
        =
        \mathbf{q}
        +
        g_{1}\,\mathbf{\Psi}^{(1)}(\mathbf{q},z)
        +
        g_{2}\,\mathbf{\Psi}^{(2)}(\mathbf{q},z)
        + \cdots
    \end{equation}
    The numerical amplitudes $g_{i}$ of the displacement fields are given below with the first- and second-order solutions.
\end{enumerate}
The \textsc{stepsic} code adopts the latter, therefore the displacement fields $\mathbf{\Psi}^{(i)}(\mathbf{q},z)$ are directly computed at the initial redshift. The velocity field is then derived by taking the time derivative of the position, yielding
\begin{align}
    \mathbf{v}^{(n)}(\mathbf{q},t)
    &=
    g_{1}\,a(z) H(z) f_{1}(z)\,\mathbf{\Psi}^{(1)}(\mathbf{q},z) \\
    &+
    g_{2}\,a(z) H(z) f_{2}(z)\,\mathbf{\Psi}^{(2)}(\mathbf{q},z)
    + \cdots\,,
\end{align}
where
\begin{equation}
    f_{i}(z)
    =
    \frac{\operatorname{d} \ln D_{i}^{({+})}(z)}{\operatorname{d} \ln a}
\end{equation}
is the logarithmic derivative of the $i$-th growth factor. For e.g., $i=1$ and $i=2$, this factor can be approximated by a power of the matter density parameter with a weak exponent calibrated from perturbation theory~\citep{Bernardeau2002} as
\begin{equation}
    f_{1}(z) \approx \Omega_{\rm m}^{\,5/9}(z)\,,
    \qquad
    f_{2}(z) \approx 2\,\Omega_{\rm m}^{\,6/11}(z)\,.
\end{equation}

\subsection{First-order LPT (Zel'dovich approximation)}
The first-order LPT solution corresponds to the Zel'dovich approximation~\citep{Zeldovich1970}, historically the earliest analytic description of the nonlinear stage of gravitational instability, in which mass elements move ballistically along the gradient of the linear gravitational potential and naturally produce the caustic, pancake-like structures observed in the cosmic web. We calculate this displacement field in Fourier space, where the Poisson equation relates the overdensity field to the gravitational potential, and the first-order displacement field is derived from that potential:
\begin{equation}
    \nabla^2 \Phi^{(1)}(\mathbf{k},z)
    =
    \delta^{(1)}(\mathbf{k},z)\,,
    \qquad
    \mathbf{\Psi}^{(1)}(\mathbf{q},z)
    =
    -\nabla \Phi^{(1)}(\mathbf{q},z)\,.
\end{equation}
In Fourier space, the second relation becomes
\begin{equation}
    \mathbf{\Psi}^{(1)}(\mathbf{k},z)
    =
    i \frac{\mathbf{k}}{k^2} \delta^{(1)}(\mathbf{k},z)\,,
\end{equation}
with the Fourier convention of Sec.~\ref{ssec:nonperiodic}, in which $\nabla\to i\mathbf{k}$.
The corresponding displacement and peculiar velocity fields are then given by
\begin{align}
    \mathbf{x}^{(1)}(\mathbf{q},z)
    &=
    \mathbf{q} + g_{1}\,\mathbf{\Psi}^{(1)}(\mathbf{q},z)\,, \label{eq:lpt1-x}
    \\
    \mathbf{v}^{(1)}(\mathbf{q},z)
    &=
    g_{1}\,a(z) H(z) f_{1}(z) \mathbf{\Psi}^{(1)}(\mathbf{q},z)\,, \label{eq:lpt1-v}
\end{align}
where $g_{1}=1$ and $\mathbf{\Psi}^{(1)}(\mathbf{q},z)$ is obtained by inverse Fourier transforming $\mathbf{\Psi}^{(1)}(\mathbf{k},z)$ back to physical space.

\subsection{Second-order LPT (tidal field corrections)}
The perturbative extension of the Zel'dovich approximation to higher orders was initiated by \citet{Moutarde1991}, who derived the second- and third-order displacement fields for $\Omega_{\rm m} = 1$. \citet{Bouchet1992} subsequently generalized the second-order solution to Friedman-Lema\^{i}tre models with arbitrary $\Omega_{\rm m}$ (with $\Lambda = 0$), and \citet{Bouchet1995} presented the complete framework including third-order results and the non-zero $\Lambda$ case. The second-order growth factor satisfies
\begin{equation} \label{eq:d2-growth}
    D_{2}^{(+)}
    \simeq
    \begin{cases}
        -\tfrac{3}{7}\,\Omega_{\rm m}^{-1/143}\,[D_{1}^{(+)}]^{2} & \text{flat, with a cosmological constant,} \\[2pt]
        -\tfrac{3}{7}\,\Omega_{\rm m}^{-2/63}\,[D_{1}^{(+)}]^{2} & \text{open, without one,}
    \end{cases}
\end{equation}
so the amplitude of the second-order displacement field,
\begin{equation} \label{eq:g2}
    g_{2}
    =
    \frac{D_{2}^{(+)}(z)}{[D_{1}^{(+)}(z)]^{2}}\,,
\end{equation}
evaluated at the initial redshift, depends only weakly on the density parameter~\citep{Bouchet1995,Bernardeau2002}. Because $\Omega_{\rm m}(z)\simeq1$ at typical starting redshifts, $g_{2}$ differs from the Einstein--de Sitter value $-3/7$ by less than $10^{-6}$ at $z=31$. For spherically symmetric perturbations, the inclusion of 2LPT improves the density contrast prediction to better than $7\%$ accuracy over the range $-0.8 \lesssim \delta \lesssim 3$~\citep{Bouchet1995}. Omitting the second-order term excites transients that decay only slowly. Simulations started from 1LPT at $z=49$ underestimate the nonlinear power spectrum by about $2\%$, $4\%$, and $8\%$ at $z=0$, 1, and 3, respectively~\citep{Crocce2006}. The 2LPT framework has since become the standard for generating accurate cosmological initial conditions, and \textsc{stepsic} optionally includes these second-order corrections. The second-order potential $\Phi^{(2)}(\mathbf{q},z)$ is sourced by the quadratic combinations of the first-order tidal tensor in physical space as
\begin{equation} \label{eq:second-order-source}
    S(\mathbf{q},z)
    \equiv
    \nabla^2 \Phi^{(2)}(\mathbf{q},z)
    =
    \sum_{i<j} \left[
        \frac{\partial^2 \Phi^{(1)}}{\partial x_i \partial x_i}
        \frac{\partial^2 \Phi^{(1)}}{\partial x_j \partial x_j}
        -
        \frac{\partial^2 \Phi^{(1)}}{\partial x_i \partial x_j}
        \frac{\partial^2 \Phi^{(1)}}{\partial x_i \partial x_j}
    \right]\,,
\end{equation}
leading to the second-order displacement field $\mathbf{\Psi}^{(2)}{=}\nabla \Phi^{(2)}$. Because $g_{2}<0$, the second-order term $g_{2}\mathbf{\Psi}^{(2)}$ in Eq.~\eqref{eq:lpt2-x} below displaces particles along $-\nabla\Phi^{(2)}$, as in \textsc{2lptic} and \textsc{monofonic}. In Fourier space, $\mathbf{\Psi}^{(2)}$ becomes
\begin{equation}
    \mathbf{\Psi}^{(2)}(\mathbf{k},z)
    =
    -i \frac{\mathbf{k}}{k^2} S(\mathbf{k},z),
\end{equation}
where $S(\mathbf{k},z)$ is the Fourier transform of the quadratic source term in Eq.~\eqref{eq:second-order-source}. The \textsc{stepsic} code computes the second derivatives of $\Phi^{(1)}$ in Fourier space and multiplies them on the mesh in real space, without de-aliasing. \textsc{monofonic} evaluates the same products with the Orszag 3/2 rule, so the two codes solve the same equations but differ numerically near the Nyquist wavenumber (Sec.~\ref{ssec:monofonic}). Using the first-order displacement and velocity fields Eqs.~\eqref{eq:lpt1-x} and \eqref{eq:lpt1-v}, the complete particle trajectories and velocities up to second order are then given by
\begin{align}
    \mathbf{x}^{(2)}(\mathbf{q},z)
    &=
    \mathbf{x}^{(1)}(\mathbf{q},z)
    +
    g_{2} \mathbf{\Psi}^{(2)}(\mathbf{q},z)\,, \label{eq:lpt2-x} \\
    \mathbf{v}^{(2)}(\mathbf{q},z)
    &=
    \mathbf{v}^{(1)}(\mathbf{q},z)
    +
    g_{2}\,a(z) H(z) f_{2}(z) \mathbf{\Psi}^{(2)}(\mathbf{q},z)\,. \label{eq:lpt2-v}
\end{align}
\section{Interpolation} \label{sec:interpolation}

The numerical scheme of our initial condition generator combines a Lagrangian particle description with Eulerian fields defined on a regular grid. Consequently, quantities computed on the mesh (e.g., displacement or velocity fields) must be interpolated to particle positions when converting between the Eulerian and Lagrangian representations.

In \textsc{stepsic}, interpolation is performed using compact-support mass assignment kernels. We employ order-$p$ B-spline assignment functions with $p \in \{1,2,3\}$, corresponding to the Nearest Grid Point (NGP), Cloud-in-Cell (CIC), and Triangular Shaped Cloud (TSC) schemes, respectively \citep{HockneyEastwood1988}. These kernels have analytically known Fourier transforms:
\begin{equation}
    W(\mathbf{k})
    =
    \prod_{i=x,y,z}
    \left[
        \frac{\sin(\pi k_i / 2k_{{\rm Ny},i})}{\pi k_i / 2k_{{\rm Ny},i}}
    \right]^{p}\,,
\end{equation}
where $k_{{\rm Ny},i}$ denotes the Nyquist frequency in the $i$-th direction. In Fourier space, the interpolation therefore acts as a multiplicative low-pass filter. As shown by \citet{Jing2005}, mass assignment modifies the measured Fourier amplitudes and introduces aliasing near the Nyquist frequency:
\begin{equation}
\begin{aligned}
    \langle |\delta_f(\mathbf{k})|^2 \rangle
    =
    &\sum_{\mathbf{n}}
    |W(\mathbf{k}+2k_{\rm Ny} \mathbf{n})|^{2} P(\mathbf{k} + 2 k_{\rm Ny} \mathbf{n}) \\
    +
    \frac{1}{N} &\sum_{\mathbf{n}} |W(\mathbf{k} + 2 k_{\rm Ny} \mathbf{n})|^{2}\,.
\end{aligned}
\end{equation}
Thus, interpolation corresponds to a controlled convolution in real space and to a known window function in Fourier space. To compensate for this smoothing when interpolating grid-based fields onto particle positions that do not coincide with grid points, we apply a deconvolution factor
\begin{equation}
    W^{-1}(\mathbf{k})
    =
    \prod_{i=x,y,z}
    \left[
        \operatorname{sinc} \left( \frac{k_i}{2 k_{{\rm Ny},i}} \right)
    \right]^{-p}\,,
\end{equation}
to the Fourier modes prior to the inverse transform. This restores the ``unsmoothed'' field amplitudes for modes below the Nyquist frequency. NGP and CIC return the node values exactly when particles lie on the grid nodes, so for them the correction is then unnecessary and overcorrects the field. TSC averages neighboring nodes and requires it even there.

\subsection{Multiresolution interpolation} \label{ssec:multires}
To achieve high resolution where needed while maintaining a large dynamic range, \textsc{stepsic} constructs a multiresolution layout by stacking several grids of increasing resolution and decreasing particle mass. This approach is motivated by the need to resolve structure formation across multiple scales simultaneously---spanning from dwarf galaxy halos requiring high mass resolution to large-scale modes that determine the cosmic web~\citep{Tormen1997,Springel2005}. The theoretical foundations for such multi-scale initial conditions have been established by \cite{HahnAbel2011}, who demonstrated that consistent multiresolution sampling preserves both large-scale power and small-scale structure in zoom-in simulations. For stereographic projections used by \textsc{StePS}, this becomes essential because the radially varying resolution naturally creates particles with different effective masses as a function of distance from the observer~\citep{Racz2018}.
 
A particle of mass $m$ represents a Lagrangian volume $m/\bar{\varrho}$ with mean spacing $\Delta(m)=(m/\bar{\varrho})^{1/3}$ and cannot sample modes shorter than about $2\Delta(m)$. Adding such modes to its displacement would only inject noise into the simulation. The \textsc{stepsic} code therefore matches the resolved wavelengths to the particle mass, following the second method of \citet[Sec.~3.2]{Racz2019}, in four steps:
\begin{enumerate}[label=(\roman*)]
    \item \textbf{Mesh hierarchy.} For each distinct particle mass $m$, we request $\lfloor L_{\rm min}/\Delta(m)\rfloor$ cells along the shortest side $L_{\rm min}$ of the enclosing box. Then, we round this request up to a multiple of the smallest even count for which every side of the box holds an even number of cells, and convert it into a mesh with cubic cells through Eq.~\eqref{eq:cubic-voxel}. Every level then spans the same box. Next, we sort the distinct masses and select $N_{\rm lev}$ of them at evenly spaced positions in the sorted list, always including the smallest and the largest. The selected masses $m_j$ and their cell sizes $\Delta_j$ define the levels $j$.
    \item \textbf{Mass bracketing.} The hierarchy generally has fewer levels than there are distinct particle masses, so the mass $m$ of a particle usually falls between the masses of two adjacent levels. For every particle, we identify these two levels: the finer one, with mass $m_{\rm f}\le m$ and the smaller cells, and the coarser one, with mass $m_{\rm c}\ge m$ and the larger cells. We then express the position of $m$ between them through the weight
    \begin{equation} \label{eq:multires-weight}
        w
        =
        \frac{m - m_{\rm f}}{m_{\rm c} - m_{\rm f}}\,,
    \end{equation}
    which grows linearly from $w=0$ at $m=m_{\rm f}$ to $w=1$ at $m=m_{\rm c}$. In the final step, $w$ is the share of the coarser level in the fields of the particle and $1-w$ the share of the finer level. A particle whose mass coincides with that of a level therefore uses that level alone. Particles lighter than the finest level or heavier than the coarsest level use that end of the hierarchy alone.
    \item \textbf{Fields per level.} We then process the levels one at a time. On level $j$ we draw the white noise from the same seed as on every other level, generate the overdensity field from it, and compute the 1LPT or 2LPT displacement and velocity fields on the mesh. Because each mode depends only on the seed and on its integer wave vector (Sec.~\ref{ssec:white-noise}), a finer level contains every mode of a coarser level with identical amplitude and phase, and adds the modes up to its own Nyquist frequency. We interpolate the fields of level $j$ to the particles it brackets at $\mathbf{q}+(\Delta_j-\Delta_{\rm min})\mathbf{1}/2$ instead of $\mathbf{q}$, where $\Delta_{\rm min}$ is the finest cell size. Each level measures the phases of its modes from its own first cell center, $-\mathbf{L}/2+\Delta_j\mathbf{1}/2$. Without this origin shift, a mode shared by two levels would lie at a different place on each. Section~\ref{ssec:open-ics} measures the effect of omitting it.
    \item \textbf{Weighted sum.} Finally, we combine the fields $\mathbf{\Psi}_{\rm f}$ and $\mathbf{\Psi}_{\rm c}$ of the two bracketing levels into the displacement of each particle,
    \begin{equation} \label{eq:multires-mixing}
        \mathbf{\Psi}
        =
        (1-w)\,\mathbf{\Psi}_{\rm f} + w\,\mathbf{\Psi}_{\rm c}\,,
    \end{equation}
    and form the same combination of the velocities.
\end{enumerate}
Equation~\eqref{eq:multires-mixing} can be rewritten as
\begin{equation} \label{eq:multires-increment}
    \mathbf{\Psi}
    =
    \mathbf{\Psi}_{\rm c}
    +
    (1-w)\,(\mathbf{\Psi}_{\rm f}-\mathbf{\Psi}_{\rm c})\,.
\end{equation}
Because the shared modes are identical, the difference $\mathbf{\Psi}_{\rm f}-\mathbf{\Psi}_{\rm c}$ consists of the modes between the two Nyquist frequencies, up to the different smoothing of the two interpolation meshes. Every particle therefore carries all modes of the coarser level and a fraction $1-w$ of the additional modes of the finer one, so the scheme acts as a low-pass filter whose cutoff moves smoothly with particle mass. All particles in the constant-mass core of radius $r_{\rm c}$ (Sec.~\ref{ssec:shells}) use the finest level. Unlike the nested sub-volumes of zoom-in generators~\citep{HahnAbel2011}, every level covers the whole enclosing box, and the cost is dominated by the finest mesh. The two approaches share the requirement that a finer level reproduces the modes of a coarser one, but meet it differently. A generator using nested grids constrains the white noise of each level with that of its parent and takes the boundary values of the Poisson equation from the parent, so its levels depend on each other. In \textsc{stepsic} the shared modes are identical by construction of the white noise and every level is solved in its own periodic box, so the levels are independent and the order in which they are processed does not affect the result.

In a nested hierarchy, however, only the coarsest level is periodic. Every finer level is an isolated sub-volume, and the waves longer than the sub-volume reach it through its parent, in the same way as the outer layers of a \textsc{StePS} domain carry the waves longer than its region of high resolution. The topology would therefore enter through the coarsest level alone (the box of the simulation in $\mathbb{T}^{3}$, and the enclosing box of Sec.~\ref{ssec:nonperiodic} for the open geometries). For a cylinder, the finer levels would in addition have to span the full length $L_z$ and remain periodic along the axis.

The choice between the two constructions is thus one between cost and exactness. Consider the cost first. The finest mesh is needed only by the lightest particles, which occupy the core and the first shells around it. In our construction this mesh nevertheless covers the whole box, so most of its cells lie where no particle uses them. Its size is set by the volume of the box, and a wider enclosing box enlarges every mesh of the hierarchy. A nested hierarchy avoids this waste. Its finest level would cover only the region that needs it, so the size of that level would be set by the region of high resolution and not by the box. A wider box would then enlarge the coarsest level alone, and the effects of a finite enclosing box could be reduced at little cost.

The price of this saving is exactness. Our levels are independent of each other, and none of them has a boundary inside the domain. A nested level, by contrast, is joined to its parent along its boundary, and this joint is not perfect. For one level of refinement, \citet[Sec.~4.3.1]{HahnAbel2011} find that the velocities agree closely with a full-resolution solution in the interior of the refined region, while in the cells at its boundary the differences are more than an order of magnitude larger. Where these boundaries lie therefore matters. In a \textsc{StePS} domain every finer level would place its boundary closer to the center, that is, closer to the region that the simulation is designed to resolve. The enclosing box acts instead on the outer layers, where the construction is coarsest by design. A \textsc{StePS}-like particle load also offers no natural place for such boundaries. Beyond $r_{\rm c}$ the particle mass increases continuously with radius, so any radius chosen for a boundary would be arbitrary, and the interpolation in mass of Eq.~\eqref{eq:multires-mixing} would still be needed to pass smoothly from one level to the next. We therefore use levels that cover the whole box. Each of them contains exactly the modes of that box. In single precision the peak memory grows by 28 bytes per mesh cell (Sec.~\ref{ssec:embedding}). A nested hierarchy becomes the better choice once the finest mesh exceeds the memory of a computing node, and is left for future work. Higher-order interpolation across levels is possible when enough levels are available and is also left for future work.

\section{Validation and tests} \label{sec:validation}

An initial condition generator is correct when, for the chosen topology, it produces the intended Gaussian realization (Sec.~\ref{sec:geometries}). Therefore, we test the periodic calculation shared by all geometries (Sects.~\ref{ssec:periodic-boxes} and \ref{ssec:monofonic}), the additional steps of the open geometries (Sects.~\ref{ssec:open-ics} and \ref{ssec:embedding}), and the initial conditions together with the \textsc{StePS} dynamics (Sec.~\ref{ssec:evolution}). The randomly generated white noise is a deterministic function of the seed and of the mode wave vector (Sec.~\ref{ssec:white-noise}), so a parameter file and its seed reproduce the initial conditions completely, and the realization can be saved for comparison with other codes.


\subsection{Periodic boxes} \label{ssec:periodic-boxes}

We begin with the periodic calculation shared by all geometries. We generate Gaussian density fields from a target $P(k)$, displace a particle load with LPT, and compare the recovered matter power spectrum with the band-averaged input spectrum on the discrete Fourier lattice. The load is a periodic glass (Sec.~\ref{ssec:glass}) of $256^{3}$ particles, relaxed with \textsc{gadget-4}~\citep{Springel2021} under reversed gravity and tiled four times along each side of the $L=1000\,h^{-1}\mathrm{Mpc}$ box, which gives $1024^{3}$ particles. Every case uses this load, so $N^{3}$ denotes the Fourier mesh and not the number of particles. Glass particles lie between the mesh nodes, where every interpolation scheme smooths the displacement field, so the interpolation window is compensated in every case (Sec.~\ref{sec:interpolation}). Particle power spectra are measured using interlaced CIC deposition with deconvolution on a $1024^{3}$ mesh, as fine as the load, so that the power of the glass beyond the Nyquist wavenumber of the Fourier mesh is not aliased into the measured range. Paired-and-fixed realizations are averaged to reduce cosmic variance~\citep{AnguloPontzen2016}.

\begin{figure}[tbp]
\centering
\includegraphics[width=\columnwidth]{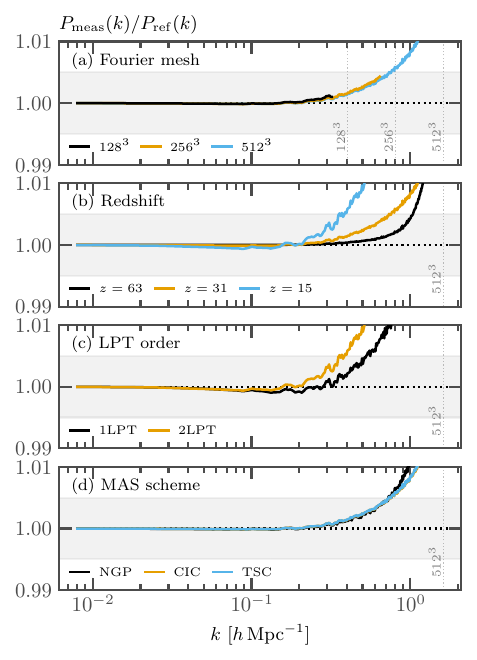}
    \caption{
        Recovery of the input matter power spectrum in a periodic box. Every panel displaces the same particle load, a $256^{3}$ glass tiled to $1024^{3}$ particles, and compensates the interpolation window. \textbf{(a)}~Fourier mesh at $z=31$ with 2LPT. \textbf{(b)}~Initial redshift with a $512^{3}$ Fourier mesh and 2LPT. \textbf{(c)}~First- and second-order LPT with a $512^{3}$ Fourier mesh at $z=15$. \textbf{(d)}~NGP, CIC, and TSC interpolation with a $512^{3}$ Fourier mesh at $z=31$ and 2LPT. The box side is $L=1000\,h^{-1}\mathrm{Mpc}$ in every panel. Shading marks a $\pm0.5\%$ deviation, and dotted lines mark the Nyquist wavenumbers of the Fourier meshes.
    }
    \label{fig:validation-grid}
\end{figure}

Fig.~\ref{fig:validation-grid} separates four numerical choices: Fourier mesh, initial redshift, LPT order, and interpolation scheme. The recovered power agrees with the input to better than $0.08\%$ at $k<0.1\,h\,\mathrm{Mpc}^{-1}$ in every case. At higher wavenumbers it rises above the input. With CIC interpolation and 2LPT at $z=31$, the excess reaches $1\%$ at $k\approx1.1\,h\,\mathrm{Mpc}^{-1}$, or $0.69\,k_{\rm Ny}$ of the $512^{3}$ mesh. Table~\ref{tab:recovery} lists the excess of every case at four wavenumbers.

The input and the measurement describe different fields. The input is the linear spectrum, whereas the measurement is the power of the displaced particles. Their density is set by the Jacobian of the mapping $\mathbf{q}\to\mathbf{q}+\boldsymbol{\Psi}$. This density equals the linear density contrast only to first order, and its higher orders couple modes of different wave vectors~\citep{Bernardeau2002}. Three contributions can therefore raise the measured power above the input, and they differ in how they depend on the mesh and on time. Errors of the field calculation, such as aliasing or imperfect compensation of the interpolation, depend on $k/k_{\rm Ny}$ and hence on the Fourier mesh. The residual power of the glass does not grow, so its share falls as the inverse square of the linear growth factor. The coupling of modes does not depend on the mesh as long as the modes that couple are resolved, and it adds terms quadratic in the linear spectrum, so its share grows as the square of the growth factor, by a factor of four each time $1+z$ halves. The measurements single out the last. At a given $k$, the three Fourier meshes agree to within $0.02$ percentage points below half of their Nyquist wavenumbers and to within $0.05$ up to $0.8\,k_{\rm Ny}$ (Table~\ref{tab:recovery}, block a), although $k/k_{\rm Ny}$ differs by up to a factor of four between them. The power of the undisplaced glass is a small part of the excess below $1\,h\,\mathrm{Mpc}^{-1}$ (Table~\ref{tab:recovery}, block e), and after its subtraction the excess at $k=0.6\,h\,\mathrm{Mpc}^{-1}$ grows by factors of $3.9$ and $4.0$ from $z=63$ to 31 and from 31 to 15 (Table~\ref{tab:recovery}, block b). Over all $k$ bins between $0.3$ and $0.8\,h\,\mathrm{Mpc}^{-1}$ the median factors are $4.00$ and $3.96$. The excess is thus the nonlinear power of the displaced particles, which the initial conditions should contain. The field calculation contributes at most the small differences between the meshes. Its amount depends on the LPT order. The second-order displacement, sourced by Eq.~\eqref{eq:second-order-source}, completes the second-order density of perturbation theory, which the Zel'dovich approximation lacks~\citep{Bernardeau2002}. This missing term is also the origin of the transients of 1LPT initial conditions~\citep{Crocce2006}. Accordingly, first-order displacements give a smaller excess than second-order ones wherever the excess is positive (Table~\ref{tab:recovery}, block c). With compensation, CIC and TSC interpolation give the same power to within $0.04\%$, and NGP gives a larger excess at high $k$ (Table~\ref{tab:recovery}, block d).

\begin{table}[tbp]
    \centering
    \begin{tabular*}{\columnwidth}{@{\extracolsep{\fill}}lcccc@{}}
        \hline\hline
        $k$ [$h\,\mathrm{Mpc}^{-1}$] & 0.1 & 0.3 & 0.6 & 1.2 \\
        \hline
        \multicolumn{5}{@{}l}{(a) Fourier mesh, $z=31$} \\
        \blockrule
        $128^{3}$ & $-0.004$ & $0.113$ & -- & -- \\
        $256^{3}$ & $-0.007$ & $0.086$ & $0.369$ & -- \\
        $512^{3}$ & $-0.007$ & $0.077$ & $0.327$ & $1.215$ \\
        \hline
        \multicolumn{5}{@{}l}{(b) Initial redshift, $512^{3}$} \\
        \blockrule
        $z=63$ & $-0.002$ & $0.019$ & $0.097$ & $1.010$ \\
        $z=31$ & $-0.007$ & $0.077$ & $0.327$ & $1.215$ \\
        $z=15$ & $-0.030$ & $0.306$ & $1.282$ & $4.054$ \\
        \hline
        \multicolumn{5}{@{}l}{(c) LPT order, $512^{3}$, $z=15$} \\
        \blockrule
        1LPT & $-0.053$ & $0.076$ & $0.611$ & $2.319$ \\
        2LPT & $-0.030$ & $0.306$ & $1.282$ & $4.054$ \\
        \hline
        \multicolumn{5}{@{}l}{(d) Interpolation, $512^{3}$, $z=31$} \\
        \blockrule
        NGP & $-0.007$ & $0.065$ & $0.288$ & $1.918$ \\
        CIC & $-0.007$ & $0.077$ & $0.327$ & $1.215$ \\
        TSC & $-0.007$ & $0.079$ & $0.331$ & $1.218$ \\
        \hline
        \multicolumn{5}{@{}l}{(e) Undisplaced glass} \\
        \blockrule
        $z=63$ & $0.000$ & $0.001$ & $0.013$ & $0.714$ \\
        $z=31$ & $0.000$ & $0.000$ & $0.003$ & $0.179$ \\
        $z=15$ & $0.000$ & $0.000$ & $0.001$ & $0.045$ \\
        \hline
    \end{tabular*}
    \caption{Excess of the recovered power over the input, $100\,(P_{\rm meas}/P_{\rm ref}-1)$, for the cases of Fig.~\ref{fig:validation-grid}. Values are in percent, at the bins nearest the stated wavenumbers. Blocks (a)--(d) correspond to the panels of Fig.~\ref{fig:validation-grid}. Block (e) gives the power of the glass before displacement relative to the input spectrum at each redshift.}
    \label{tab:recovery}
\end{table}

\begin{figure}[tbp]
\centering
\includegraphics[width=\columnwidth]{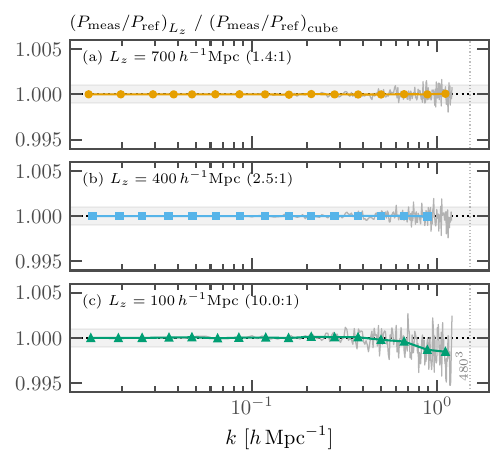}
    \caption{
        Power recovery in periodic slabs relative to the cubic box as the periodic length $L_z$ is reduced from $1000$ to $100\,h^{-1}\mathrm{Mpc}$ at fixed transverse size $1000\,h^{-1}\mathrm{Mpc}$. Each panel shows, for one slab with $L_z=700$ \textbf{(a)}, $400$ \textbf{(b)}, and $100\,h^{-1}\mathrm{Mpc}$ \textbf{(c)}, the ratio $P_{\rm meas}/P_{\rm ref}$ of the measured particle power to the band-averaged input power on the discrete modes of that box, divided by the same ratio for the cube. Gray lines use the native $k$ bins, and symbols average $P_{\rm meas}$ and $P_{\rm ref}$ over logarithmic bins of width $\Delta\log_{10}k=1/8$, weighted by mode count. The cell size $\Delta=L_{z,\min}/N_{\rm input}=2.08\,h^{-1}\mathrm{Mpc}$ is held constant, so the meshes differ in shape but not in spatial resolution. The measurement extends to $0.8\,k_{\rm Ny}$ with $k_{\rm Ny}=\pi/\Delta$, marked by dotted lines in the panels. All realizations use 2LPT at $z=31$, CIC interpolation, and paired-and-fixed averaging. Shading marks a $\pm0.1\%$ deviation.
    }
    \label{fig:squish-validation}
\end{figure}

The same recovery test can be applied to a periodic slab. Fig.~\ref{fig:squish-validation} keeps the physical cell size fixed while shortening one dimension from $1000$ to $100\,h^{-1}\mathrm{Mpc}$, and divides the recovered power of each slab by that of the cube. Table~\ref{tab:squish} lists the largest deviations of this ratio from unity. Averaged over logarithmic bins, they stay within the shaded $\pm0.1\%$ band except for the $10{:}1$ slab above $0.8\,h\,\mathrm{Mpc}^{-1}$, where the recovered power of the cube itself exceeds the input by 4--7\%. Individual $k$ bins scatter more. This scatter scales approximately as the inverse square root of the number of modes per bin, which is proportional to the slab volume. The represented modes therefore retain the expected amplitudes. This result tests the normalization and numerical treatment of those modes, but it does not show that a thin periodic slab has the same statistics as a region of equal shape cut from an isotropic three-dimensional field. It should not, because a periodic slab is a $\mathbb{T}^{3}$ topology in its own right (Sec.~\ref{sec:geometries}), and the task of the generator is to realize the modes of that topology with the correct amplitudes, which is what Fig.~\ref{fig:squish-validation} tests. A periodic cube likewise lacks modes longer than its side and is anisotropic on the largest scales~\citep{Racz2021}, and a slab makes this anisotropy stronger.

\begin{table}[tbp]
    \centering
    \begin{tabular*}{\columnwidth}{@{\extracolsep{\fill}}lcccc@{}}
        \hline\hline
        & & \multicolumn{2}{c}{Logarithmic bins} & Native bins \\
        $L_z$ [$h^{-1}\mathrm{Mpc}$] & $L/L_z$ & $k<0.6$ & $k\geq0.6$ & all $k$ \\
        \hline
        700 & 1.4 & 0.004 & 0.007 & 0.20 \\
        400 & 2.5 & 0.003 & 0.004 & 0.24 \\
        100 & 10 & 0.020 & 0.15 & 0.52 \\
        \hline
    \end{tabular*}
    \caption{Largest deviation from unity of the slab-to-cube power recovery ratio of Fig.~\ref{fig:squish-validation}, in percent. Wavenumbers are in $h\,\mathrm{Mpc}^{-1}$ and extend to $0.8\,k_{\rm Ny}=1.21\,h\,\mathrm{Mpc}^{-1}$. In the logarithmic bins at $k\geq0.6\,h\,\mathrm{Mpc}^{-1}$ the ratio of the $L_z=100\,h^{-1}\mathrm{Mpc}$ slab lies below unity.}
    \label{tab:squish}
\end{table}

\begin{figure}[tbp]
\centering
\includegraphics[width=\columnwidth]{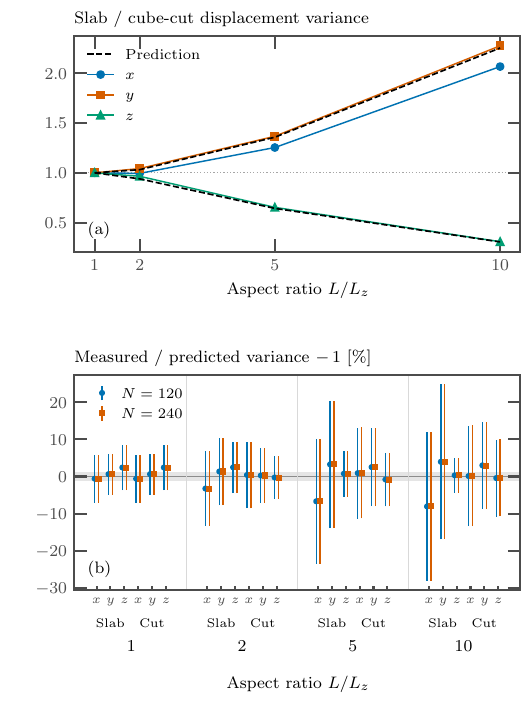}
    \caption{
        Displacement variance in periodic slabs and in equally sized regions cut from a periodic cube of side $L=1000\,h^{-1}\mathrm{Mpc}$. \textbf{(a)}~Ratio of slab to cube-cut variance of the displacement components $x$, $y$ (transverse), and $z$ (along the short axis) as a function of $L/L_z$. Symbols are ensemble means at transverse resolution $N=240$, and dashed curves are discrete-mode predictions. \textbf{(b)}~Fractional difference between each measured variance and its prediction for $N=120$ and $240$. Error bars are simultaneous 95\% confidence intervals over all displayed comparisons. The calculation uses 20 Gaussian realizations, first-order LPT at $z=31$, and a Gaussian density filter with $k_{\rm s}=0.1\,h\,\mathrm{Mpc}^{-1}$.
    }
    \label{fig:slab-covariance}
\end{figure}

The difference between the two models is apparent in the displacement field. Fig.~\ref{fig:slab-covariance} compares a periodic slab with an equally sized region cut from a periodic cube. As the slab becomes thinner, it loses long-wavelength axial modes and gains transverse variance relative to the cube cut. At $L/L_z=10$, the predicted slab-to-cut variance ratio is approximately $2.25$ for the transverse components and $0.31$ for the axial component. The measurements reproduce this directional trend, and every simultaneous 95\% interval includes the prediction, but intervals of up to $\pm21\%$ cannot test it at the percent level. The same consideration applies to the periodic axis of a cylindrical domain, whose period is usually much shorter than its diameter.

\subsection{Comparison with \textsc{monofonic}} \label{ssec:monofonic}

As an external benchmark, we generated matched periodic initial conditions with the publicly available \textsc{monofonic} code~\citep{Hahn2021} using identical cosmological parameters, box size, and resolution. Because \textsc{monofonic} implements a different random-number generator, the white noise field was generated in \textsc{stepsic} and supplied to \textsc{monofonic} through its constraint-field interface. The \textsc{camb}-generated linear power spectrum from \textsc{stepsic} was also supplied as an external transfer function, so both calculations used the same input spectrum. \textsc{monofonic} divides an imported field by its sample standard deviation. We disabled this step because the saved \textsc{stepsic} field has zero Nyquist planes, so that both codes use identical mode amplitudes.

\begin{figure}[tbp]
\centering
\includegraphics[width=\columnwidth]{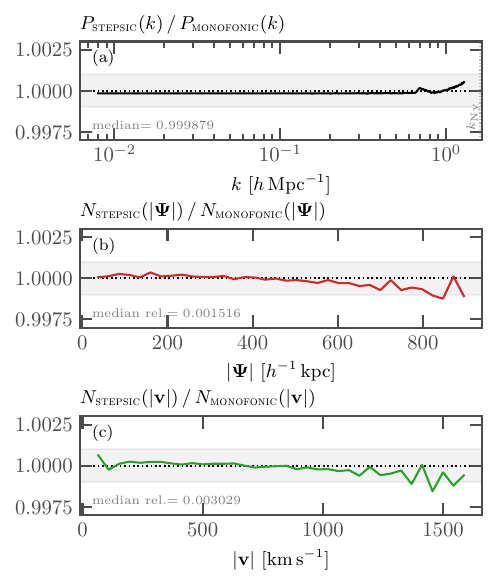}
    \caption{
        Periodic comparison between \textsc{stepsic} and \textsc{monofonic} for the same white noise realization, input spectrum, cosmology, and $512^3$-particle box of side $1000\,h^{-1}\mathrm{Mpc}$ at $z=31$. Both calculations use 2LPT. \textbf{(a)}~Matter power-spectrum ratio. \textbf{(b)}~Ratio of displacement-magnitude histograms. \textbf{(c)}~Ratio of velocity-magnitude histograms. The values printed in panels (b) and (c) are median per-particle relative residuals. Shading marks a $\pm0.1\%$ deviation from unity.
    }
    \label{fig:validation-monofonic}
\end{figure}

Fig.~\ref{fig:validation-monofonic} compares the particle power spectrum and the distributions of displacement and velocity magnitudes. Table~\ref{tab:monofonic} lists the per-particle agreement. The power-spectrum ratio stays between $0.99983$ and $1.00052$ at $k\leq0.8\,k_{\rm Ny}$. It is constant at $0.9998$ for $k<0.1\,h\,\mathrm{Mpc}^{-1}$, an offset consistent with the different numerical integrals with which the two codes normalize the shared spectrum to $\sigma_8$. The rise toward the Nyquist wavenumber comes from the second-order term. The magnitude distributions are insensitive to the second-order correction at leading order, because its cross-correlation with the first-order field vanishes for a Gaussian field. With first-order LPT, the displacements of the two codes differ by a relative root-mean-square (RMS) of $7.8\times10^{-5}$ on $64^3$, $128^3$, and $256^3$ lattices, the amplitude offset of the normalization, so the larger per-particle residuals of the 2LPT comparison come from the second-order term. The velocity residual is larger than the displacement residual because this term enters the velocity with a growth rate about twice that of the first-order term.

\begin{figure}[tbp]
\centering
\includegraphics[width=\columnwidth]{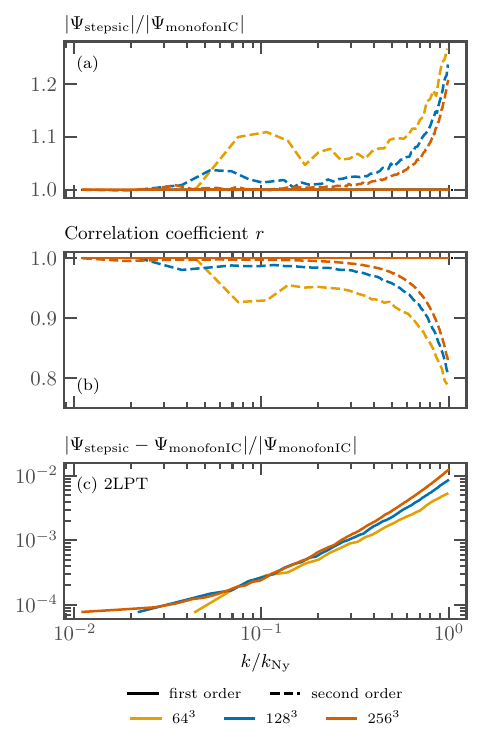}
    \caption{
        Mode-by-mode comparison between \textsc{stepsic} and \textsc{monofonic} of the first-order displacement (solid) and of the isolated second-order correction $g_{2}\mathbf{\Psi}^{(2)}$ (dashed), the difference between 2LPT and 1LPT positions from the same white noise, on $64^3$, $128^3$, and $256^3$ lattices in a box of side $1000\,h^{-1}\mathrm{Mpc}$ at $z=31$. \textbf{(a)}~Amplitude ratio of the modes in each Fourier shell. \textbf{(b)}~Correlation coefficient of the modes in each shell. \textbf{(c)}~RMS difference of the total 2LPT displacements of the two codes in each shell, relative to the \textsc{monofonic} displacement. The first-order curves lie at the normalization offset, $0.99992$, in (a) and at unity in (b).
    }
    \label{fig:isolated-orders}
\end{figure}

We therefore also compare the second-order correction directly. The difference between 2LPT and 1LPT positions generated from the same white noise isolates $g_{2}\mathbf{\Psi}^{(2)}$ in each code. Fig.~\ref{fig:isolated-orders} compares the Fourier modes of these fields, and those of the first-order displacements, on $64^3$, $128^3$, and $256^3$ lattices. The first-order displacements agree mode by mode at every wavenumber, with the constant amplitude ratio of the normalization and a correlation coefficient of unity. The isolated second-order correction has the same sign in both codes. Toward the Nyquist wavenumber its amplitude ratio rises above unity and its correlation coefficient falls, on every lattice, and at lower wavenumbers the difference shrinks as the lattice is refined. A difference of normalization or growth rate would be the same at every wavenumber and on every lattice. Aliasing in the quadratic source depends on the lattice. \textsc{monofonic} evaluates the source with the Orszag 3/2 rule, whereas \textsc{stepsic}, like \textsc{2lptic}~\citep{Crocce2006}, multiplies the derivative fields directly on the mesh, so that products of modes near the Nyquist wavenumber fold back onto lower wavenumbers. Built with its aliased convolver, \textsc{monofonic} reproduces the \textsc{stepsic} correction in every shell on all three lattices, with a correlation coefficient that differs from unity by less than $2\times10^{-9}$. The amplitude ratio is then the square of the normalization offset of the first order, because the correction is quadratic in the first-order field.

The differences in panels (a) and (b) are fractions of the second-order correction, which at $z=31$ is a small part of the displacement. On the $256^3$ lattice its RMS is $0.006$ of the first-order RMS. Panel (c) shows what they change in the total displacement. At large scales the two codes differ by the normalization offset of the first order. Toward the Nyquist wavenumber, where the second-order correction is largest, the difference grows, and it is this difference that sets the per-particle residuals of Table~\ref{tab:monofonic}. \citet{Michaux2021} found the aliasing of the 2LPT source likewise confined to the highest wavenumbers. In their simulations it evolved as a decaying transient and changed the power spectrum at $z\lesssim3$ by an amount of order $10^{-3}$, for a start later than ours chosen to maximize it.

\begin{table}[tbp]
    \centering
    \begin{tabular*}{\columnwidth}{@{\extracolsep{\fill}}lcc@{}}
        \hline\hline
        & Displacement & Velocity \\
        \hline
        Median residual [\%] & $0.15$ & $0.30$ \\
        99th-percentile residual [\%] & $0.88$ & $1.8$ \\
        Largest histogram deviation [\%] & $0.12$ & $0.16$ \\
        \hline
    \end{tabular*}
    \caption{Per-particle agreement between \textsc{stepsic} and \textsc{monofonic} in the $512^3$ 2LPT comparison of Fig.~\ref{fig:validation-monofonic}. The residual of a particle is the magnitude of the difference between the vectors of the two codes, relative to the magnitude of the \textsc{monofonic} vector. The histogram deviation is the difference from unity of the ratio of the magnitude histograms, over the bins that hold at least $1\%$ of the peak count.}
    \label{tab:monofonic}
\end{table}

\subsection{Spherical and cylindrical initial conditions} \label{ssec:open-ics}

Evolved simulations started from the spherical and cylindrical constructions that \textsc{stepsic} implements have been published, and the spherical ones have been compared with standard $\mathbb{T}^{3}$ simulations. \citet{Racz2018} evolved a spherical \textsc{StePS} simulation and a \textsc{gadget-2} simulation~\citep{Springel2005Gadget2} of the enclosing periodic box from the same \textsc{2lptic} realization and found close agreement between the angular power spectra of the two runs in radial shells at 35--65 and 285--315\,Mpc. \citet{Racz2019} initialized a \textsc{StePS} simulation with the cosmology and random seed of the Millennium simulation~\citep{Springel2005} and compared its density field at $z=0$ directly with the compactified Millennium snapshot. The structures coincide in the central region, and the differences at large radii arise because compactification and time evolution do not commute. Both spherical simulations were initialized with \texttt{StePS\_IC}, which computed the displacement fields with \textsc{2lptic}/\textsc{n-genic} (Sec.~\ref{sec:intro}). For the cylinder, the power spectra of the simulation of \citet{Racz2026}, initialized with a pre-release version of \textsc{stepsic}, follow the linear and nonlinear predictions. These are tests of evolved simulations. The \textsc{stepsic} code keeps the construction that they validate and performs the work of the external generators itself, and the comparisons with $\mathbb{T}^{3}$ simulations extend to its initial conditions because each step of its calculation is tested against a reference. The periodic calculation recovers the input spectrum (Sec.~\ref{ssec:periodic-boxes}) and agrees with \textsc{monofonic} for the same white noise (Sec.~\ref{ssec:monofonic}). The steps that \textsc{stepsic} adds to it can be compared with an exact reference in the initial conditions themselves, for both open geometries together: the interpolation to relaxed glasses and the consistency of the multiresolution levels (this section) and the dependence on the enclosing periodic box (Sec.~\ref{ssec:embedding}).

\begin{figure*}[t]
\centering
\includegraphics[width=\textwidth]{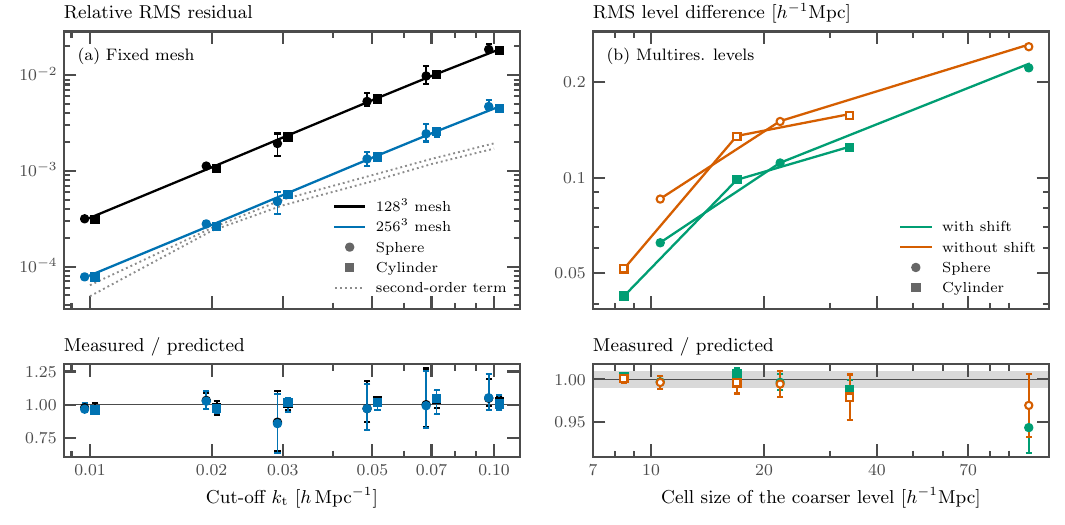}
    \caption{
        Interpolation of the displacement field to the particles of the relaxed spherical (circles) and cylindrical (squares) glasses at $z=31$. Symbols are measurements and solid lines are sums over modes. \textbf{(a)}~Relative RMS difference $\epsilon$, Eq.~\eqref{eq:rms-residual}, between the 2LPT displacements interpolated from a fixed $128^3$ or $256^3$ mesh and the direct sum of the same Fourier modes at the same particles, against the cutoff $k_{\rm t}$ of the input spectrum. Every symbol uses the same 1792 particles, 256 in each of seven radial intervals, and its bar spans the values of the seven intervals. The lines are the error $\epsilon_{\rm CIC}$ of CIC interpolation at uniformly distributed positions, Eq.~\eqref{eq:cic-residual}, and the lower panel shows $\epsilon/\epsilon_{\rm CIC}$. Dotted lines give the RMS of the second-order term relative to that of the displacement. \textbf{(b)}~RMS difference between the 1LPT displacements that two adjacent levels of the multiresolution test hierarchies of Table~\ref{tab:domains} give the same particle, against the cell size of the coarser level, with (filled) and without (open) the origin shift of Sec.~\ref{ssec:multires}. Symbols are means over 16 realizations with their standard errors, and the lines are $\sigma_{\rm fc}$, Eq.~\eqref{eq:level-difference}. The lower panel shows the ratio of the measured variance to $\sigma_{\rm fc}^2$, and its shading marks $\pm1\%$.
    }
    \label{fig:glass-interpolation}
\end{figure*}

On a glass, the fields are sampled at irregular positions in a coordinate frame centered on the domain. We compare the production fixed-mesh particle displacements on relaxed spherical and cylindrical glasses with a separate Fourier calculation of the retained modes. The reference uses the saved density field and forms the quadratic 2LPT source by explicit convolution, without calling the production LPT solver or its interpolation routine. The cylindrical glass is that of Table~\ref{tab:domains}, and the spherical one is a larger glass of 34,914,653 particles. Both are scaled to a radius of $250\,h^{-1}\mathrm{Mpc}$ in a Fourier cube of side $1000\,h^{-1}\mathrm{Mpc}$, fifteen copies of the scaled cylindrical glass fill the axial period, and the fields are interpolated with CIC without compensation. The input spectrum is cut off smoothly at a wavenumber $k_{\rm t}$, and the direct sum retains the modes up to $1.5k_{\rm t}$. Modes above make a negligible contribution. The quadratic source then extends to $3k_{\rm t}$, at most $0.30\,h\,\mathrm{Mpc}^{-1}$, below the Nyquist wavenumber of the coarser mesh, $0.40\,h\,\mathrm{Mpc}^{-1}$, so both meshes represent it without aliasing. The resolution dependence of the field calculation itself is tested in Sec.~\ref{ssec:periodic-boxes}.

For a measured displacement $\mathbf{u}$ and its reference $\mathbf{u}_{\rm ref}$, the relative RMS residual is
\begin{equation} \label{eq:rms-residual}
    \epsilon = \left[\frac{\sum|\mathbf{u}-\mathbf{u}_{\rm ref}|^2}{\sum|\mathbf{u}_{\rm ref}|^2}\right]^{1/2},
\end{equation}
with sums over the selected particles. If the mesh holds the field exactly at its cell centers and every particle is read at its own position on the mesh, the only contribution to $\epsilon$ is the linear interpolation between cells. At positions distributed uniformly within a cell of size $\Delta$, CIC interpolation returns a mode of wave vector $\mathbf{k}$ with mean amplitude $\prod_i\mathrm{sinc}^2(k_i\Delta/2)$ and mean square amplitude $\prod_i(2+\cos k_i\Delta)/3$, where $\mathrm{sinc}\,x=\sin x/x$, and different modes do not correlate. The expected residual is therefore
\begin{equation} \label{eq:cic-residual}
    \epsilon_{\rm CIC}^2 = \frac{\sum_{\mathbf{k}}|\mathbf{\Psi}_{\mathbf{k}}|^2\left[\prod_i\frac{2+\cos k_i\Delta}{3}-2\prod_i\mathrm{sinc}^2\frac{k_i\Delta}{2}+1\right]}{\sum_{\mathbf{k}}|\mathbf{\Psi}_{\mathbf{k}}|^2},
\end{equation}
where $\mathbf{\Psi}_{\mathbf{k}}$ are the Fourier coefficients of the reference displacement.

Fig.~\ref{fig:glass-interpolation}(a) compares $\epsilon$ with $\epsilon_{\rm CIC}$. The residual grows with the cutoff, and doubling the mesh resolution reduces it by a factor of four at every cutoff, as Eq.~\eqref{eq:cic-residual} predicts for these smooth fields. The ratio $\epsilon/\epsilon_{\rm CIC}$ lies between $0.96$ and $1.04$ for the cylinder in all twelve cases, and between $0.86$ and $1.05$ for the sphere, whose radial intervals scatter more (bars in the lower panel). The comparison is sensitive to where the particles are read on the mesh. Evaluating the cylindrical $128^3$ field at $k_{\rm t}=0.03\,h\,\mathrm{Mpc}^{-1}$ half a cell away from the particle positions raises $\epsilon$ from $0.225\%$ to $7.1\%$. It is less sensitive to the second-order term, whose RMS (dotted lines) is a fraction of the residual on the $128^3$ mesh and approaches it on the $256^3$ mesh. Removing the term from the reference raises $\epsilon$ for the cylinder at $256^3$ and $k_{\rm t}=0.03\,h\,\mathrm{Mpc}^{-1}$ from $0.057\%$ to $0.077\%$. The comparison thus confirms the presence and the approximate amplitude of the second-order term at the glass positions. Its accuracy is tested mode by mode in Sec.~\ref{ssec:monofonic}. Because the reference shares the saved density field, the normalization of the Gaussian field is tested separately, in Sec.~\ref{ssec:periodic-boxes}.

These fixed-mesh measurements test interpolation at the glass positions. The multiresolution scheme adds one requirement (Sec.~\ref{ssec:multires}). A finer level must contain the modes of the adjacent coarser level unchanged, so that the difference between the two levels' displacements at the same particles consists only of the modes that the finer level adds. We evaluated every level of the four-level test hierarchies of Table~\ref{tab:domains} at every particle of the spherical and cylindrical glasses of that table, as the production code does, with 1LPT fields of fixed amplitude for 16 seeds. If the shared modes are identical, the variance of the difference $\mathbf{\Psi}_{\rm f}-\mathbf{\Psi}_{\rm c}$ is
\begin{equation} \label{eq:level-difference}
\begin{aligned}
    \sigma_{\rm fc}^2
    = {} & \sum_{\mathbf{k}\in K_{\rm f}}|\mathbf{\Psi}_{\mathbf{k}}|^2\prod_i\frac{2+\cos k_i\Delta_{\rm f}}{3} \\
    & + \sum_{\mathbf{k}\in K_{\rm c}}|\mathbf{\Psi}_{\mathbf{k}}|^2\prod_i\frac{2+\cos k_i\Delta_{\rm c}}{3} \\
    & - 2\sum_{\mathbf{k}\in K_{\rm c}}|\mathbf{\Psi}_{\mathbf{k}}|^2\,\mathrm{Re}\,\big\langle t_{\rm f}(\mathbf{k})\,t_{\rm c}^{*}(\mathbf{k})\big\rangle\,,
\end{aligned}
\end{equation}
where $|\mathbf{\Psi}_{\mathbf{k}}|^2=P(k)/(Vk^2)$ is the term of mode $\mathbf{k}$ in the trace of the 1LPT displacement covariance, Eq.~\eqref{eq:periodic-covariance}, which the fixed amplitudes give every realization exactly, $\Delta_{\rm f}$ and $\Delta_{\rm c}$ are the cell sizes of the finer and the coarser level, and $K_{\rm f}$ and $K_{\rm c}$ are the sets of modes below their Nyquist frequencies, so that $K_{\rm c}\subset K_{\rm f}$. The factor
\begin{equation*}
    t_j(\mathbf{k}) = \prod_i\left[(1-u_{j,i})\,e^{-ik_iu_{j,i}\Delta_j}+u_{j,i}\,e^{ik_i(1-u_{j,i})\Delta_j}\right]
\end{equation*}
is the CIC transfer of a mode on level $j$: the interpolated value at the point where the particle reads that level, divided by the value of the level's mode at that point. The point lies a fraction $u_{j,i}$ of a cell beyond the preceding mesh node along axis $i$. The average runs over positions distributed uniformly in the box, and different modes do not correlate. The mean square of $t_j$ is the factor $\prod_i(2+\cos k_i\Delta_j)/3$ of Eq.~\eqref{eq:cic-residual}, so the first two terms are the variances of the two levels, and the last term is the covariance of their shared modes. Without the origin shift, the shared modes on the two levels are offset by $(\Delta_{\rm c}-\Delta_{\rm f})/2$ along each axis, which multiplies $t_{\rm f}t_{\rm c}^{*}$ by $\prod_i\exp[ik_i(\Delta_{\rm c}-\Delta_{\rm f})/2]$. Fig.~\ref{fig:glass-interpolation}(b) shows the RMS differences. For five of the six pairs of levels, the measured variance agrees with $\sigma_{\rm fc}^2$ to within $1.2\%$. For the coarsest spherical pair, levels 1 and 2, the ratio is $0.94\pm0.03$. Evaluating both levels at $\mathbf{q}$, without the origin shift of step (iii), increases the RMS difference by 17--37\%, as Eq.~\eqref{eq:level-difference} predicts.

\begin{table}[tbp]
    \centering
    \begin{tabular*}{\columnwidth}{@{\extracolsep{\fill}}lcc@{}}
        \hline\hline
        & Sphere & Cylinder \\
        \hline
        \multicolumn{3}{@{}l}{Glass} \\
        \blockrule
        Particles & 3,374,078 & 1,983,884 \\
        Radius $R$ & 508 & 508 \\
        Axial period $L_z$ & -- & 135 \\
        High-resolution radius & 105 & 106 \\
        \hline
        \multicolumn{3}{@{}l}{Cell size $\Delta_j$ of level $j$} \\
        \blockrule
        $j=1$ & 102 & 33.9 \\
        $j=2$ & 22.1 & 16.9 \\
        $j=3$ & 10.6 & 8.46 \\
        $j=4$ & 6.77 & 6.77 \\
        \hline
    \end{tabular*}
    \caption{Domains of the tests of the open geometries in Sects.~\ref{ssec:open-ics}--\ref{ssec:evolution}. Lengths are in $h^{-1}\mathrm{Mpc}$. The test hierarchies have four levels, coarser than the automatic choice, and are used for Fig.~\ref{fig:glass-interpolation}(b) and for the test of the complete path in Sec.~\ref{ssec:embedding}. The spherical glass, relaxed with $R=531\,\mathrm{Mpc}$, is stretched to the radius of the cylindrical one (Sec.~\ref{ssec:glass}). The evolved simulations of Sec.~\ref{ssec:evolution} use the cylindrical glass with $R=750\,\mathrm{Mpc}$ and $L_z=200\,\mathrm{Mpc}$. Particles with at most eight times the smallest particle mass lie within the high-resolution radius of the center of the sphere or the axis of the cylinder.}
    \label{tab:domains}
\end{table}

\subsection{Dependence on the enclosing periodic box} \label{ssec:embedding}

The initial fields of the open geometries are computed in a periodic box whose side is $\alpha$ times the domain diameter (Sec.~\ref{ssec:nonperiodic}). We compare boxes with $\alpha=1$, 2, and 3 with the infinite Universe, which for the sphere is the three-dimensional continuum of wave vectors and for the cylinder a continuum of transverse wave vectors with the axial wavenumbers $2\pi n/L_z$ of the topology. The domains are those of Table~\ref{tab:domains}. All values are for $z=31$.

\begin{figure*}[t]
\centering
\includegraphics[width=\textwidth]{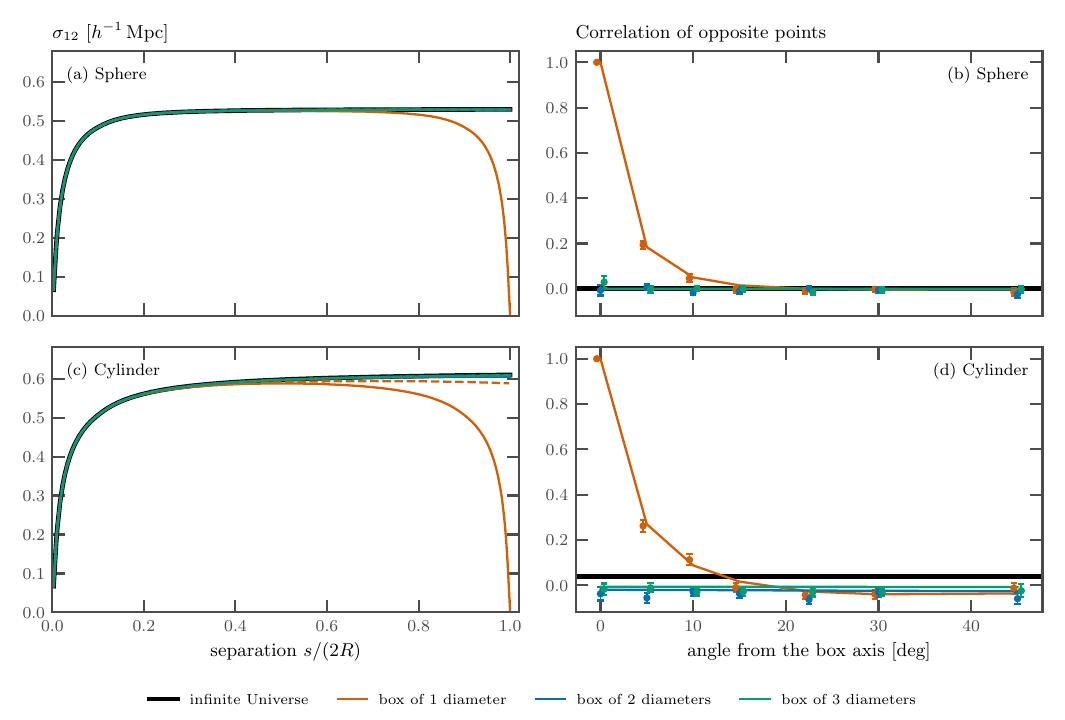}
    \caption{
        Relative displacement of two points and correlation of the displacements at opposite points of the boundary at $z=31$, in the infinite Universe and in enclosing boxes of one, two, and three domain diameters. \textbf{(a)}~RMS relative displacement $\sigma_{12}$ in the sphere against the separation $s$ in units of the diameter $2R$, along a box axis (solid) and along a body diagonal of the box (dashed). \textbf{(b)}~Correlation coefficient $C(2R\hat{\mathbf{n}})/C(\mathbf{0})$ of the displacements at two opposite points of the surface of the sphere, against the angle between the line that connects them and the box axis. \textbf{(c)}~As (a) for the cylinder, with the separation perpendicular to its axis, along a transverse box axis (solid) and a transverse diagonal (dashed). \textbf{(d)}~Same as (b) but for the lateral surface of the cylinder. Lines are sums over the modes of each box, and points are 200 generated 2LPT realizations. Only the box of one diameter departs visibly from the infinite Universe, at separations that approach the diameter and within eight to ten degrees of a box axis.
    }
    \label{fig:embedding-two-point}
\end{figure*}

For a Gaussian field the two-point statistics are a complete description, and in a periodic box they depend only on the separation vector of the two points (Sec.~\ref{ssec:nonperiodic}). We therefore evaluate the mean square relative displacement of two points separated by $\mathbf{s}$,
\begin{equation} \label{eq:relative-displacement}
    \sigma_{12}^{2}(\mathbf{s})
    =
    \left\langle\left|\boldsymbol\Psi(\mathbf{q}+\mathbf{s})-\boldsymbol\Psi(\mathbf{q})\right|^2\right\rangle
    =
    2\left[C(\mathbf{0})-C(\mathbf{s})\right]\,,
\end{equation}
where $C$ is the trace of the displacement covariance of Eq.~\eqref{eq:periodic-covariance}, as a sum over the modes of each box and as an integral for the infinite Universe. No random field enters. The sums use meshes with cells of $2.8\,h^{-1}\mathrm{Mpc}$ in both geometries and the filter $|F|^2=\exp(-k^2/k_{\rm s}^2)$ with $k_{\rm s}=0.4\,h\,\mathrm{Mpc}^{-1}$, which removes the modes near the Nyquist frequency of these meshes. The differences between a box and the infinite Universe come from long waves and do not depend on the filter.

Fig.~\ref{fig:embedding-two-point}(a) and (c) show $\sigma_{12}$, and Table~\ref{tab:embedding-two-point} lists its values. The box changes $\sigma_{12}$ least at short separations, which contain the region for which the simulation is designed. Over all separations between particles of the region of high resolution (Table~\ref{tab:domains}), the largest difference between the box with $\alpha=1$ and the infinite Universe is $0.0002\,h^{-1}\mathrm{Mpc}$ for the sphere and $0.0014\,h^{-1}\mathrm{Mpc}$ for the cylinder. The differences grow beyond the domain radius $R$ along a box axis, and at $s=2R$ along a box axis $\sigma_{12}$ vanishes, because the two points are the same point of the periodic box. Along a diagonal of the box the differences stay small (dashed lines). With $\alpha=2$ and $\alpha=3$, $\sigma_{12}$ follows the infinite Universe to within $0.005\,h^{-1}\mathrm{Mpc}$ at every separation inside the domain.

Panels (b) and (d) show the same effect at the boundary, through the correlation coefficient $C(2R\hat{\mathbf{n}})/C(\mathbf{0})$ of the displacements at two opposite points, listed in Table~\ref{tab:embedding-opposite}. With $\alpha=1$ it is unity on a box axis and falls within eight to ten degrees of it, because through the periodic image the opposite point is a neighbor at distance $4R\sin(\theta/2)$. The coefficient exceeds $0.1$ on $3\%$ of the surface of the sphere and on $21\%$ of the lateral surface of the cylinder. With $\alpha=2$ and 3 it stays close to zero in every direction. The cylinder, whose long waves are two-dimensional, converges more slowly.

\begin{table}[tbp]
    \centering
    \begin{tabular*}{\columnwidth}{@{\extracolsep{\fill}}lcccc@{}}
        \hline\hline
        $s$ & Infinite & $\alpha=1$ & $\alpha=2$ & $\alpha=3$ \\
        \hline
        \multicolumn{5}{@{}l}{Sphere} \\
        \blockrule
        $100\,h^{-1}\mathrm{Mpc}$ & $0.4825$ & $0.4825$ & $0.4825$ & $0.4825$ \\
        $R$ & $0.5282$ & $0.5268$ & $0.5282$ & $0.5282$ \\
        $1.5R$ & $0.5296$ & $0.5213$ & $0.5294$ & $0.5295$ \\
        $2R$ & $0.5300$ & $0$ & $0.5297$ & $0.5300$ \\
        \hline
        \multicolumn{5}{@{}l}{Cylinder} \\
        \blockrule
        $100\,h^{-1}\mathrm{Mpc}$ & $0.5041$ & $0.5038$ & $0.5040$ & $0.5041$ \\
        $R$ & $0.5973$ & $0.5878$ & $0.5963$ & $0.5970$ \\
        $1.5R$ & $0.6055$ & $0.5709$ & $0.6032$ & $0.6049$ \\
        $2R$ & $0.6096$ & $0$ & $0.6049$ & $0.6084$ \\
        \hline
    \end{tabular*}
    \caption{RMS relative displacement $\sigma_{12}$ of two points separated by $s$ along a box axis at $z=31$, in the infinite Universe and in enclosing boxes of one, two, and three domain diameters. Values are in $h^{-1}\mathrm{Mpc}$ and are sums over the modes of each box, and integrals for the infinite Universe (Eq.~\eqref{eq:relative-displacement}). $R$ is the domain radius (Table~\ref{tab:domains}). The first row is the separation on the mesh nearest to $100\,h^{-1}\mathrm{Mpc}$, $99.9\,h^{-1}\mathrm{Mpc}$ for the sphere and $98.7\,h^{-1}\mathrm{Mpc}$ for the cylinder. With $\alpha=1$, two points separated by $2R$ along a box axis are the same point of the periodic box.}
    \label{tab:embedding-two-point}
\end{table}

\begin{table}[tbp]
    \centering
    \begin{tabular*}{\columnwidth}{@{\extracolsep{\fill}}lcccc@{}}
        \hline\hline
        $\theta$ & Infinite & $\alpha=1$ & $\alpha=2$ & $\alpha=3$ \\
        \hline
        \multicolumn{5}{@{}l}{Sphere} \\
        \blockrule
        $0^\circ$ & $0.002$ & $1.000$ & $-0.001$ & $0.000$ \\
        $5^\circ$ & $0.002$ & $0.184$ & $-0.001$ & $0.000$ \\
        $10^\circ$ & $0.002$ & $0.050$ & $-0.001$ & $0.000$ \\
        $15^\circ$ & $0.002$ & $0.015$ & $-0.001$ & $0.000$ \\
        \hline
        \multicolumn{5}{@{}l}{Cylinder} \\
        \blockrule
        $0^\circ$ & $0.039$ & $1.000$ & $-0.020$ & $-0.007$ \\
        $5^\circ$ & $0.039$ & $0.269$ & $-0.020$ & $-0.007$ \\
        $10^\circ$ & $0.039$ & $0.088$ & $-0.021$ & $-0.007$ \\
        $15^\circ$ & $0.039$ & $0.016$ & $-0.022$ & $-0.007$ \\
        \hline
    \end{tabular*}
    \caption{Correlation coefficient $C(2R\hat{\mathbf{n}})/C(\mathbf{0})$ of the displacements at two opposite boundary points at $z=31$, for directions $\hat{\mathbf{n}}$ at an angle $\theta$ from a box axis. Values are sums over the modes of each box, and integrals for the infinite Universe, which is isotropic. The direction $\hat{\mathbf{n}}$ is rotated from a box axis toward a face diagonal for the sphere and lies in the plane perpendicular to the axis for the cylinder.}
    \label{tab:embedding-opposite}
\end{table}

\begin{table}[tbp]
    \centering
    \footnotesize
    \setlength{\tabcolsep}{2.5pt}
    \begin{tabular*}{\columnwidth}{@{\extracolsep{\fill}}lcccccc@{}}
        \hline\hline
        & \multicolumn{2}{c}{$\sigma(\bar{\delta}_V)$} & \multicolumn{2}{c}{RMS of $\bar{T}_{ij}$} & \multicolumn{2}{c}{Off-diagonal share} \\
        Box & Pred. & Meas. & Pred. & Meas. & Pred. & Meas. \\
        \hline
        \multicolumn{7}{@{}l}{Sphere} \\
        \blockrule
        Infinite & 2.03 & -- & 1.66 & -- & 0.60 & -- \\
        $\alpha=1$ & 1.57 & $1.55\pm0.07$ & 1.28 & $1.32\pm0.04$ & 0.24 & $0.25\pm0.02$ \\
        $\alpha=2$ & 2.01 & $2.01\pm0.10$ & 1.64 & $1.65\pm0.04$ & 0.55 & $0.54\pm0.02$ \\
        $\alpha=3$ & 2.02 & $2.01\pm0.08$ & 1.65 & $1.62\pm0.04$ & 0.59 & $0.56\pm0.02$ \\
        \hline
        \multicolumn{7}{@{}l}{Cylinder} \\
        \blockrule
        Infinite & 4.06 & -- & 2.87 & -- & 0.50 & -- \\
        $\alpha=1$ & 2.35 & $2.29\pm0.11$ & 1.66 & $1.74\pm0.07$ & 0.34 & $0.30\pm0.03$ \\
        $\alpha=2$ & 3.95 & $4.02\pm0.17$ & 2.79 & $2.83\pm0.10$ & 0.40 & $0.33\pm0.03$ \\
        $\alpha=3$ & 4.03 & $4.14\pm0.19$ & 2.85 & $2.91\pm0.11$ & 0.47 & $0.46\pm0.04$ \\
        \hline
    \end{tabular*}
    \caption{Mean density contrast and mean shear of the domain at $z=31$ in the infinite Universe and in enclosing boxes of one, two, and three domain diameters. Standard deviation of the mean density contrast $\bar{\delta}_V$ (Eq.~\eqref{eq:monopole}) and RMS of the mean shear $\bar{T}_{ij}$ (Eq.~\eqref{eq:mean-strain}), both in units of $10^{-4}$, and the share of the variance of $\bar{T}_{ij}$ in its off-diagonal components in the frame of the box. Predicted values are sums over the modes of each box, and integrals for the infinite Universe. Measured values are from 200 Gaussian realizations of the 2LPT displacements at points on the boundary of the domain, on meshes with cells of about $9\,h^{-1}\mathrm{Mpc}$. Uncertainties are standard deviations over bootstrap resamples of the realizations.}
    \label{tab:embedding}
\end{table}

The box also sets properties of the domain as a whole: the mean density contrast $\bar{\delta}_V$ of Eq.~\eqref{eq:monopole} and the mean shear $\bar{T}_{ij}$, the traceless part of Eq.~\eqref{eq:mean-strain}. For a Gaussian field their variances are sums over the modes of the box,
\begin{equation} \label{eq:domain-variance}
    \begin{aligned}
        \sigma^{2}(\bar{\delta}_V)
        &=
        \frac{1}{V_{\rm box}}\sum_{\mathbf{k}}P(k)\,|\hat{\chi}_V(\mathbf{k})|^2\,, \\
        \sigma^{2}(\bar{T}_{ij})
        &=
        \frac{1}{V_{\rm box}}\sum_{\mathbf{k}}P(k)\,|\hat{\chi}_V(\mathbf{k})|^2
        \left(\hat{k}_i\hat{k}_j-\frac{\delta_{ij}}{d}\right)^{2},
    \end{aligned}
\end{equation}
and integrals for the infinite Universe. Here $V_{\rm box}$ is the volume of the box, $\hat{\chi}_V$ the Fourier transform of the domain indicator normalized to unit volume average, and $d=3$ for the sphere and $d=2$ for the cylinder. For the cylinder only modes with $k_z=0$ contribute to an average over the full period, and $\hat{\mathbf{k}}$ is their transverse direction. Summed over $i$ and $j$, the weight of the shear is $1-1/d$ for every mode, so the RMS of $\bar{T}_{ij}$ is $(1-1/d)^{1/2}\,\sigma(\bar{\delta}_V)$ in every box. The box changes the size of the mean shear together with the mean density contrast, and its direction through the share of the off-diagonal components. Table~\ref{tab:embedding} lists the three quantities. The measurements use 200 realizations of the 2LPT displacements at points on the boundary of the domain, through the boundary integrals of Eqs.~\eqref{eq:monopole} and \eqref{eq:mean-strain}. They agree with the predictions within three standard errors, and the largest difference is the off-diagonal share of the cylinder with $\alpha=2$. In every realization the boundary integrals agree with the same averages taken directly from the density modes to within $5\%$ of the predicted standard deviation, the difference being CIC interpolation error. The same realizations give the points in Fig.~\ref{fig:embedding-two-point}(b) and (d). With $\alpha=1$, the standard deviation of $\bar{\delta}_V$ is smaller than in the infinite Universe for both domains, and with it the mean shear, which is also aligned with the box axes. Its off-diagonal components carry a smaller share of its variance. These are properties of an ensemble. A single realization has a valid mean density and mean shear, drawn from a narrower distribution. With $\alpha=2$ the predictions approach those of the infinite Universe, the off-diagonal share of the cylinder most slowly. For the whole periodic box the mean density contrast vanishes by construction.

As a check of the complete path, we also displaced both glasses with the multiresolution scheme of Sec.~\ref{ssec:multires} in the three boxes, with the test hierarchies of Table~\ref{tab:domains}, for 20 seeds and their sign pairs, and measured $\sigma_{12}^{2}$ over the same 65,536 mass-weighted particle pairs in each separation bin. The changes of $\sigma_{12}^{2}$ between the boxes agree with the sums over modes within the scatter of the seeds, and the scatter is larger than the changes. For pairs of the sphere at $s\simeq447\,h^{-1}\mathrm{Mpc}$, the sums give an increase of $0.68\times10^{-3}\,(h^{-1}\mathrm{Mpc})^2$ from $\alpha=1$ to $\alpha=2$, and the ensemble gives $(2.4\pm2.0)\times10^{-3}$. The values quoted above are therefore those of the sums over modes.

A wider box does not change the resolution. In the multiresolution mode the cell of each mesh follows from the particle mass (Sec.~\ref{ssec:multires}), so every mesh grows with the box, by $\alpha^3$ for the sphere and by $\alpha^2$ for the cylinder. We timed the command line on the spherical and cylindrical glasses of Table~\ref{tab:domains} with a fixed cell in the three boxes, on meshes of up to $3\times10^{8}$ cells, in single precision on four threads. The peak memory grows by 28 bytes and the wall time by about 190\,ns per mesh cell. For the spherical glass the automatic hierarchy would place the finest mesh at $1022^3$ cells with $\alpha=1$ and $2042^3$ with $\alpha=2$. We did not run these meshes. Extrapolated, they need a peak memory of 28 and 222\,GiB.

We keep $\alpha=1$ as the default, for three reasons. First, \textsc{StePS} is a zoom construction, in which the mass resolution is highest at the center and decreases outward, and the outer layers provide the large-scale environment of the central region. The enclosing box acts where this construction is coarsest by design. In the region of high resolution the initial conditions have the two-point statistics of the infinite Universe to within $0.0014\,h^{-1}\mathrm{Mpc}$ in $\sigma_{12}$. The differences begin at separations above the domain radius and are largest at the boundary. Second, on the scale of the domain the box narrows the distribution from which the mean density and the mean shear of a realization are drawn, but it does not make any single realization invalid. Third, all \textsc{StePS} simulations to date were started from boxes with $\alpha=1$~\citep[e.g.,][]{Racz2018,Racz2019,Racz2026}. The measurements above concern the initial conditions. For one cylindrical realization evolved to $z=0$ from boxes with $\alpha=1$ and $\alpha=2$, the power spectra of the region of high resolution differ by at most $1.05\%$ in any Fourier shell once the shift common to the region is removed (Sec.~\ref{ssec:evolution}). A wider box is the better choice when the domain as a whole is the object of study: for ensembles in which the scatter of the mean density or the mean shear between realizations matters, and for statistics at separations comparable to the domain radius, such as large-angle statistics of the outer shells. The box is an input parameter, and $\alpha=2$ removes the differences listed above at eight times the mesh cells for the sphere and four times for the cylinder.

\subsection{Evolution of matched initial conditions} \label{ssec:evolution}

\begin{figure}[tbp]
\centering
\includegraphics[width=\columnwidth]{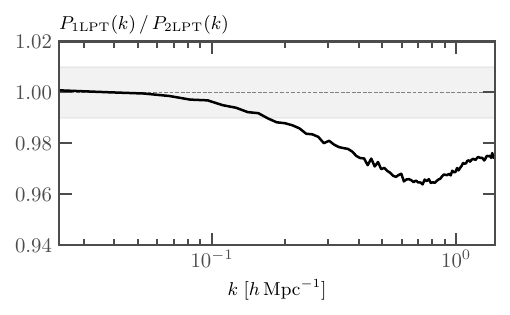}
    \caption{
        Ratio of the $z=0$ matter power spectra from matched cylindrical \textsc{StePS} simulations initialized at $z=31$ with 1LPT and 2LPT, on the cylindrical glass of Table~\ref{tab:domains} with $D_{\rm 4D}=35\,\mathrm{Mpc}$. Both spectra were measured with the FKP estimator of \citet{Racz2026} using three radial bins and a $512^3$ mesh, with shot noise subtracted. The curve ends at $1.45\,h\,\mathrm{Mpc}^{-1}$, below the wavenumber at which the estimator switches between radial bins. Shading marks a $\pm1\%$ deviation from unity.
    }
    \label{fig:validation-cylindrical}
\end{figure}

The initial conditions and the \textsc{StePS} dynamics can be tested together through the transients that 1LPT initial conditions excite (Sec.~\ref{sec:lpt}). We evolved two matched cylindrical $S^{1}{\times}\mathbb{R}^{2}$ realizations from $z=31$ to $z=0$, one started from 1LPT and one from 2LPT, with the same white noise field and the relaxed cylindrical glass of Table~\ref{tab:domains}. Fig.~\ref{fig:cylindrical-density} shows the projected density field of the 2LPT run. The final power spectra were measured with the non-periodic Feldman--Kaiser--Peacock (FKP) estimator of \citet{Racz2026}, with the same random catalog and settings for both runs. Fig.~\ref{fig:validation-cylindrical} shows their ratio. On large scales the runs agree, and $P_{\rm 1LPT}/P_{\rm 2LPT}$ is within $0.1\%$ of unity at $k<0.06\,h\,\mathrm{Mpc}^{-1}$. At smaller scales the 1LPT run has less power, and the ratio reaches its minimum, $0.964$, at $k\approx0.7\,h\,\mathrm{Mpc}^{-1}$. This is the deficit that the transients of 1LPT initial conditions produce (Sec.~\ref{sec:lpt}). It is somewhat larger than the $2\%$ that \citet{Crocce2006} found at $z=0$ for a start at $z=49$, as expected for our later start, which leaves the transients less time to decay.

A second matched pair tests the enclosing box of Sec.~\ref{ssec:embedding} after the evolution. We evolved one more realization twice with the glass and the settings above, once from initial conditions computed in a box with $\alpha=1$ and once with $\alpha=2$. On the finest mesh, the white noise of the wider box was matched inside the domain to that of the first run and was independent outside it. The coarser levels take their modes from this field. The two initial conditions then differ by what the wider box adds. Table~\ref{tab:evolved-pair} lists the differences between the displacements of the same particle in the two runs. In the initial conditions they consist of a shift common to the region of high resolution (Table~\ref{tab:domains}), a much smaller scatter about it inside that region, and a larger one in the outermost layer. All three grow during the evolution.

The shift is part of a displacement of the domain as a whole, which the long waves of the wider box add. In the mass-weighted sum over the particles the pairwise forces cancel, so the transverse center of mass of the domain feels only the force with which \textsc{StePS} represents the homogeneous exterior of the cylinder, and the difference $X$ between the two runs obeys the same linear equation. This equation and that of the linear growth factor,
\begin{equation} \label{eq:center-of-mass}
    \ddot{X} + 2H\dot{X} = 2\pi G\bar{\varrho}\,X\,,
    \qquad
    \ddot{D}_{1}^{(+)} + 2H\dot{D}_{1}^{(+)} = 4\pi G\bar{\varrho}\,D_{1}^{(+)}\,,
\end{equation}
where $X$ is comoving and $\bar{\varrho}$ is the physical mean matter density at the time, differ by a factor of two in the source term, so the displacement of the domain grows more slowly than a wave of the infinite Universe. From $z=31$ to $z=0$ it grew from $0.056$ to $0.54\,h^{-1}\mathrm{Mpc}$, by a factor of $9.7$ where $D_{1}^{(+)}$ grows by $25.1$, and it follows the solution of Eq.~\eqref{eq:center-of-mass} to within $0.4\%$.

\begin{figure}[tbp]
\centering
\includegraphics[width=\columnwidth]{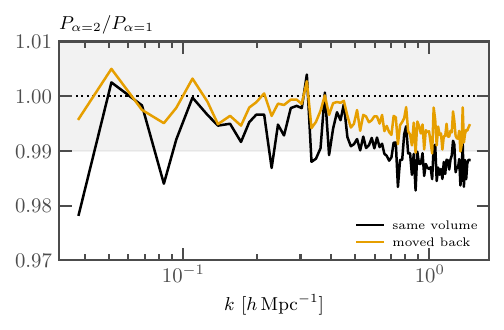}
    \caption{
        Ratio of the $z=0$ power spectra of the region of high resolution, within $106\,h^{-1}\mathrm{Mpc}$ of the axis, in two \textsc{StePS} simulations of one cylindrical realization started at $z=31$ from enclosing boxes of two ($\alpha=2$) and one ($\alpha=1$) domain diameters, with the same white noise inside the domain. Both spectra were measured with the FKP estimator of \citet{Racz2026} on a $512^3$ mesh, in the same 97 Fourier shells from $0.038$ to $1.46\,h\,\mathrm{Mpc}^{-1}$. Black: the same volume in both runs. Orange: the particles of the $\alpha=2$ run moved back by the displacement common to the region before the selection. Shading marks a $\pm1\%$ deviation from unity.
    }
    \label{fig:evolved-pair}
\end{figure}

At $z=0$ we measured the power spectrum of the region with the FKP estimator, in the same volume and the same Fourier shells for both runs (Fig.~\ref{fig:evolved-pair}). By then the shift common to the region has grown, so the fixed volume does not hold the same matter in the two runs. We therefore repeated the measurement with the particles of the second run moved back by this shift before the selection. The two spectra then agree to within $1.05\%$ in every Fourier shell. The run with $\alpha=2$ has slightly less power, and its region is less dense, with a mean density contrast of $2.9\times10^{-2}$ against $3.2\times10^{-2}$. This is a single realization of a single geometry, and the mean density contrast of its domain differs between the two boxes by a third to a half of the RMS difference expected from Table~\ref{tab:embedding}.

\begin{table}[tbp]
    \centering
    \begin{tabular*}{\columnwidth}{@{\extracolsep{\fill}}lcc@{}}
        \hline\hline
        & $z=31$ & $z=0$ \\
        \hline
        Common shift of the region & 0.050 & 0.39 \\
        RMS in the region about the shift & 0.0048 & 0.37 \\
        RMS in the outermost layer about its mean & 0.19 & 2.5 \\
        \hline
    \end{tabular*}
    \caption{Difference between the displacements of the same particles in the two evolutions of Fig.~\ref{fig:evolved-pair}, in $h^{-1}\mathrm{Mpc}$. The region is the region of high resolution (Table~\ref{tab:domains}), and the outermost layer lies between $0.93R$ and $R$ from the axis. The common shift is the magnitude of the mean difference in the region.}
    \label{tab:evolved-pair}
\end{table}

\section{Computational performance}\label{sec:performance}

We measured the wall time and peak resident-set size (RSS) of complete 1LPT and 2LPT calculations with $N_{\rm grid}=N_{\rm mesh}\equiv N=64$, 128, 256, and 512. The timed interval extends from process launch through the completed HDF5 snapshot and therefore includes the \textsc{camb} calculation, construction of the LPT fields, particle interpolation, and file output. All configurations use single precision, grid particles, and uncompensated CIC interpolation in a periodic cube of side $1000\,h^{-1}\mathrm{Mpc}$ at $z=31$. Each configuration was launched in a separate process. We report the median and full range of three runs.

\begin{figure}
  \centering
  \includegraphics[width=\columnwidth]{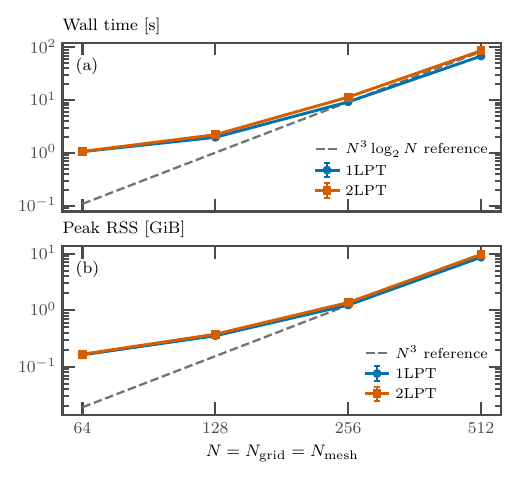}
  \caption{
    End-to-end wall time \textbf{(a)} and peak resident-set size \textbf{(b)} for periodic grid initial conditions with $N^3$ particles on an $N^3$ mesh. Points show medians and error bars span the full range of three independent process launches. All ranges are smaller than the markers. The dashed curves show $N^3\log_2 N$ scaling in (a) and $N^3$ scaling in (b), normalized to the 1LPT measurement at $N=256$, and they are guides rather than fits. Measurements were performed on one Intel Core i7-13700KF processor with 16 physical cores and 24 hardware threads, all of which were available to the Fourier transforms.}
  \label{fig:computational-performance}
\end{figure}

The largest calculation contains $512^3\simeq1.34\times10^8$ particles. Its median wall time is $68.6\,\mathrm{s}$ for 1LPT and $86.5\,\mathrm{s}$ for 2LPT, with peak RSS values of $8.66$ and $9.66\,\mathrm{GiB}$, respectively. Across all four resolutions, adding the second-order displacement increases the median runtime by $1{-}26\%$ and peak RSS by $3{-}12\%$. The absolute timings apply to the benchmark system, whereas these ratios compare the two LPT orders under the same conditions.

Between $N=256$ and $512$, the wall time increases by factors of 7.4 for 1LPT and 7.6 for 2LPT, compared with a factor of 9 for the $N^3\log_2 N$ guide. Peak RSS increases by factors of $7.03$ and $7.07$ over the same interval, close to the factor of 8 expected for quantities stored on a three-dimensional mesh or for $N^3$ particles. The implementation limits additional working memory by constructing spectral temporaries in slabs, reusing Fourier buffers, processing field components one at a time, and interpolating particles in chunks. Below $N=256$ both curves flatten because a cost independent of $N$ dominates. Of the $1.1\,\mathrm{s}$ needed at $N=64$, importing the libraries takes about $0.3\,\mathrm{s}$ and the \textsc{camb} calculation about $0.6\,\mathrm{s}$, and these two steps alone reach $0.13$ of the $0.16{-}0.17\,\mathrm{GiB}$ peak RSS. At $N=512$ the total CPU time is only 1.2 (1LPT) and 1.3 (2LPT) times the wall time, so most of the runtime is spent in single-threaded operations rather than in the multithreaded Fourier transforms. The benchmark uses one machine and one thread count and therefore does not measure parallel efficiency. It characterizes the regular periodic workload. Costs for spherical and cylindrical multiresolution calculations additionally depend on their particle counts and mesh hierarchies.

We implement \textsc{stepsic} in Python to integrate with the modern scientific Python ecosystem and to keep the code more accessible for inspection and extension. It retains interpreter-level orchestration and offers less direct control over low-level optimization and distributed-memory parallelization than a dedicated compiled implementation such as \textsc{monofonic}. The dominant compute-heavy operations, however, execute in compiled routines provided by \textsc{numpy} and \textsc{scipy}, while the Python interface to \textsc{camb} invokes its compiled solver. These limitations must be weighed against support for the rectangular cuboid, spherical, and cylindrical topologies described in this work, for which conventional periodic initial condition generators are not applicable.

The benchmark measures the complete program and includes neither component-level profiling nor a comparison with a functionally equivalent compiled implementation, so it cannot quantify a slowdown attributable specifically to Python. The relevant practical measure is therefore the end-to-end cost. Initial conditions are generated once per simulation, and the complete $512^3$ calculation takes less than one and a half minutes with less than $10\,\mathrm{GiB}$ of memory on the benchmark workstation. The current implementation is practical for single-node workloads at the tested scale. Meshes that exceed the memory of one node would require a distributed-memory implementation. For ensembles that share a cosmology, supplying a precomputed power spectrum avoids repeating the \textsc{camb} calculation and reduces the setup cost of each realization.

\section{Summary and outlook} \label{sec:summary}
We have presented \textsc{stepsic}, an open-source initial condition generator for cosmological $N$-body simulations in the periodic cuboid, spherical, and cylindrical topologies supported by the \textsc{StePS} code. Each topology is a choice with its own advantages and limitations, and the generator produces, for the chosen topology, a Gaussian realization with the statistics of that topology. It is the stand-alone successor of \texttt{StePS\_IC}, which initialized the earlier spherical \textsc{StePS} simulations with displacement fields from external periodic generators. These simulations were compared with periodic simulations started from the same realizations, among them the Millennium simulation. The \textsc{stepsic} code keeps that construction and computes the fields itself. An earlier version of it initialized the cylindrical simulations, and periodic cuboids with arbitrary aspect ratios are new. The package implements Lagrangian perturbation theory up to second order, supports cuboid, spherical, and cylindrical particle loads, and provides a multiresolution scheme that matches the resolved wavelengths to the local particle mass. Standard and paired-and-fixed Gaussian realizations are supported, and the code can either compute the linear matter power spectrum internally via \textsc{camb} or accept a user-supplied spectrum. For prepared spherical and cylindrical glasses, particle displacements differ from an independent sum over the retained Fourier modes by the error expected for linear interpolation between mesh cells, between $0.86$ and $1.05$ of its predicted value. The fields of the open geometries are computed in a periodic box of one domain diameter by default. In the high-resolution region the RMS relative displacement of two points then differs from its value in the infinite Universe by at most $0.0014\,h^{-1}\mathrm{Mpc}$ out of $0.5\,h^{-1}\mathrm{Mpc}$. The differences are confined to separations above the domain radius and to the spread of the mean density and mean shear of the domain between realizations, and a box of two diameters removes them at eight (sphere) or four (cylinder) times the mesh cells. In one cylindrical realization evolved to $z=0$ from boxes of one and of two diameters, the high-resolution region is displaced as a whole by $0.39\,h^{-1}\mathrm{Mpc}$ between the runs. With this displacement removed, its power spectra differ by at most $1.05\%$. For the same white noise, the periodic calculation, which takes the place of the external generators of \texttt{StePS\_IC}, agrees with \textsc{monofonic} to within $0.05\%$ in the power spectrum and $0.15\%$ in the median particle displacement. Evolved to $z=0$ in a cylindrical domain, first-order initial conditions leave up to $3.6\%$ less nonlinear power than second-order ones, as expected from their transients.

Several extensions are planned for future releases. On the perturbation theory side, a third-order LPT implementation would further reduce transient errors at moderate initial redshifts with minimal additional computational cost and is a reasonable future extension of our initial condition generator. Support for primordial non-Gaussianity, through local-type $f_{\rm NL}$ modifications of the initial density field, would broaden the range of cosmological models that can be tested.


\begin{acknowledgements}
  This work was supported by the Hungarian Ministry of Innovation and Technology NRDI Office grant OTKA NN147550 and NKKP-153428, the KDP-2021 program from the source of the NRDI fund, and by the Hungarian National Research, Development and Innovation Office (NKFIH) under Contract No. 2025-1.1.5-NEMZ\_KI-2025-0005. GR acknowledges the support of the Research Council of Finland grant 354905 and the support by the European Research Council via ERC Consolidator grant KETJU No. 818930. IS acknowledges NASA ROSES grants 80NSSC24K1489 and 24-ADAP24-0074, contract number 80NM0018F0610 via a JPL sub-award. The authors wish to acknowledge the CSC -- IT Center for Science, Finland, and the Wigner Scientific Computing Laboratory (WSCLAB), Hungary, for computational resources.
\end{acknowledgements}

\bibliographystyle{aa}
\bibliography{bibliography}

\end{document}